\documentclass[letterpaper,10pt,journal,final,twocolumn]{IEEEtran}

 \usepackage{cite}
\usepackage{amssymb,amsmath,color}
\usepackage[english]{babel}
\usepackage{graphicx,epstopdf,float,tikz}
\usetikzlibrary{arrows,positioning,calc}

  \usepackage{hyperref}
  \hypersetup{
  	colorlinks=true,
  	linkcolor=black,
  	filecolor=black,      
  	urlcolor=blue,
  	citecolor=blue
  }
  
 \newcommand{\QED}{$\square$}
 
\newtheorem{lemma}{Lemma}
\newtheorem{proposition}{Proposition}
\newtheorem{remark}{Remark}
\newtheorem{theorem}{Theorem}

\newtheorem{definition}{Definition}

\newcounter{assc}
\newenvironment{ass}{%
	\unskip\par\addvspace{.5em}\noindent%
	\refstepcounter{assc}%
	\textbf{A\theassc)}\itshape%
}{\par\addvspace{.5em}\ignorespaces} %

\newcommand{\R}{\mathbb R}
\newcommand{\B}{\mathbb B}

\newcommand{\N}{\mathbb N}

\newcommand{\Rplus}{\R_+}

\newcommand{\Nz}{\N^*}

\newcommand{\cA}{\mathcal A}
\newcommand{\cB}{\mathcal B}
\newcommand{\cC}{\mathcal C}
\newcommand{\cD}{\mathcal D}
\newcommand{\cE}{\mathcal E}

\newcommand{\cG}{\mathcal G}
\newcommand{\cH}{\mathcal H}
\newcommand{\cI}{\mathcal I}
\newcommand{\cJ}{\mathcal J}
\newcommand{\cK}{\mathcal K}
\newcommand{\cL}{\mathcal L}
\newcommand{\cM}{\mathcal M}

\newcommand{\cO}{\mathcal O}

\newcommand{\cT}{\mathcal T}
\newcommand{\cU}{\mathcal U}

\newcommand{\cX}{\mathcal X}

\newcommand{\rC}{\mathrm{C}}
\newcommand{\rD}{\mathrm{D}}

\newcommand{\rM}{\mathrm{M}}

\newcommand{\rS}{\mathrm{S}}

\newcommand{\rW}{\mathrm{W}}
\newcommand{\rX}{\mathrm{X}}

\newcommand{\rZ}{\mathrm{Z}}

\newcommand{\x}{\times}
\newcommand{\st}{\,\mid\,}
\newcommand{\setto}{\rightrightarrows}

\newcommand{\und}{\underline}
\newcommand{\inv}{^{-1}}
\newcommand{\pinv}{^\dagger} 
\newcommand{\closed}{\overline}
\newcommand{\tim}{\varsigma}

\DeclareMathOperator{\col}{col}
\DeclareMathOperator{\sat}{sat}

\DeclareMathOperator{\diag}{diag}

\DeclareMathOperator{\esssup}{ess.sup}
\DeclareMathOperator{\dom}{dom}

\DeclareMathOperator{\msv}{msv}

\DeclareMathOperator{\jumps}{dom_j}
\DeclareMathOperator{\argminOp}{argmin}

\newcommand{\ball}{\B}

\newcommand{\xb}{\mathbf{x}}
\newcommand{\sr}{^\star}
\newcommand{\argmin}[1]{\underset{#1}{\argminOp}\,}
\newcommand{\Xb}{\mathbf{X}}
\newcommand{\pb}{{p}}

\newcommand{\y}{y}
\newcommand{\e}{e}

\newcommand{\vhi}{\varphi}
\newcommand{\Thmap}{\vartheta}

\newcommand{\upT}{\overline{\rm T}}
\newcommand{\unT}{\underline{\rm T}}
\newcommand{\rr}{{r}}

\newcommand{\dz}{{n_z}}
\newcommand{\dx}{{n_x}} 
\newcommand{\dy}{{n_{\y}}}
\newcommand{\du}{{n_u}}
\newcommand{\deta}{{n_\eta}}

\newcommand{\dth}{{n_\theta}}
\newcommand{\dw}{{n_w}}

\newcommand{\fb}{\mathbf{f}}
\newcommand{\bb}{\mathbf{b}}
\newcommand{\sol}{\phi}
\newcommand{\placeholder}{\star}

\newcommand{\pd}[2]{\dfrac{\partial #1}{\partial #2}}

\newcommand{\satlev}{{\rm M}}

\newcommand{\win}{\alpha_{\rm in}}
\newcommand{\wout}{\alpha_{\rm out}}
\newcommand{\vep}{\varepsilon}
\newcommand{\costf}{{\rm g}}
\newcommand{\regf}{{\rho}}
\DeclareMathOperator{\optmap}{Opt}

\newcommand{\din}{d_{\rm in}}
\newcommand{\dout}{d_{\rm out}}

\newcommand{\ee}{{\rm e}}

\newcommand{\qedenv}{\hfill$\triangleleft$}

\def\u{\frak{u}}

\newcommand{\Ae}{\overline{A}}
\newcommand{\Be}{\overline{B}}

\newcommand{\SPD}{\mathbb{SPD}}
\newcommand{\Sn}{\mathbb{S}}

\catcode`\@=11
\def\downparenfill{$\m@th\braceld\leaders\vrule\hfill\bracerd$}
\def\overparen#1{\mathop{\vbox{\ialign{##\crcr\crcr \noalign{\kern0.4ex}
				\downparenfill\crcr\noalign{\kern0.4ex\nointerlineskip}
				$\hfil\displaystyle{#1}\hfil$\crcr}}}\limits}
\def\dotoverparen#1{\mathop{\vbox{\ialign{##\crcr\hfill$\cdot $\hfill\crcr 
				\noalign{\kern-2.3ex}
				\downparenfill\crcr\noalign{\kern0.4ex\nointerlineskip}
				$\hfil\displaystyle{#1}\hfil$\crcr}}}\limits}
\catcode`\@=12

\allowdisplaybreaks
\def\middlebreak {\nulldelimiterspace0pt
	\allowbreak\mskip 0mu plus .5mu \nulldelimiterspace0pt}%

\title{
Approximate Nonlinear Regulation via Identification-Based Adaptive Internal Models
}
\author{Michelangelo Bin, Pauline Bernard, and Lorenzo Marconi%
	\thanks{%
		M. Bin is with the Dept. of Electrical and Electronic Engineering, Imperial College London, UK. P. Bernard   is with the Centre Automatique et Syst\`{e}mes, MINES ParisTech, PSL University, Paris, France. L. Marconi   is with the CASY-DEI, University of Bologna, Italy.    
	}
}

\begin{document}

\onecolumn 
\vspace{4em}
\begin{quote}
	This is the post peer-review accepted manuscript of: M. Bin, P. Bernard and L. Marconi, ``Approximate Nonlinear Regulation via Identification-Based Adaptive Internal Models,'' accepted for publication in IEEE Transaction on Automatic Control. The published version is available online at: \url{https://doi.org/10.1109/TAC.2020.3020563}
\end{quote}

\vspace{5em}
\begin{quote}
	\emph{\textcopyright{}~2020 IEEE.  Personal use of this material is permitted.  Permission from IEEE must be obtained for all other uses, in any current or future media, including reprinting/republishing this material for advertising or promotional purposes, creating new collective works, for resale or redistribution to servers or lists, or reuse of any copyrighted component of this work in other works.}
\end{quote}

\twocolumn

\maketitle

\begin{abstract}
	This paper concerns the problem of adaptive output regulation for multivariable nonlinear systems in normal form. We present a regulator employing an adaptive internal model of the exogenous signals based on the theory of nonlinear Luenberger observers. Adaptation is performed by means of  discrete-time system identification schemes,  in which  every algorithm fulfilling some optimality and stability conditions can be used. Practical and approximate regulation results are given relating the prediction capabilities of the identified model to the asymptotic bound on the regulated variables, which become asymptotic whenever a ``right'' internal model exists in the identifier's model set. The proposed approach, moreover, does not require ``high-gain'' stabilization actions. 
\end{abstract}

\section{Introduction} 
In this paper we consider the problem of adaptive output regulation for multivariable nonlinear systems of the form 
\begin{equation}\label{s:plant}
\begin{array}{lcl} 
\dot{z}&=& f(w,z,x) \\ 
\dot{x}&=& A x + B \big(q(w,z,x) + b(w,z,x) u\big)\\
\y &=& C x,
\end{array} 
\end{equation}
in which $(z,x)\in\R^\dz\x \R^\dx$ is the state of the plant, $u$ and $\y$, both taking values in $\R^\dy$, are the control input and the measured output,  $w\in\R^\dw$ is an exogenous input, $f:\R^\dw\x\R^\dz\x\R^\dx\to \R^\dz$, $q:\R^\dw\x\R^\dz\x\R^\dx\to\R^\dy$ and $b:\R^\dw\x\R^\dz\x\R^\dx\to\R^{\dy\x\dy}$ are continuous functions and, for some $\rr\in\N$, $A$ $B$ and $C$ are matrices defined as
\begin{align*}
A&:=\begin{pmatrix}
0_{\rr\dy\x\dy}\vline & \begin{matrix}
I_{(\rr-1)\dy}\\ 0_{\dy\x(\rr-1)\dy)}
\end{matrix} 
\end{pmatrix}, & B&:=\begin{pmatrix}
0_{(\rr-1)\dy\x \dy }\\ I_\dy 
\end{pmatrix}, \\ C&:=\begin{pmatrix}
I_\dy & 0_{\dy\x(\rr-1)\dy}
\end{pmatrix}.
\end{align*}
Namely, $\dx=\rr \dy$ and $x$ is a chain of $\rr$ integrators of dimension $\dy$.
The output regulation problem associated to system~\eqref{s:plant} consists in finding an output-feedback controller that (i) ensures boundedness of the closed-loop trajectories whenever $w$ is bounded, and (ii) asymptotically removes the effect of $w$ on the regulated output $\y$, thus ideally obtaining $\y(t) \to 0$ as $t\to\infty$. Output regulation is representative of many problems of practical interest depending on the role played by the exogenous signal $w$. For instance, simple \emph{stabilization} is obtained when $w$ is not present, \emph{disturbance rejection} is achieved when $w$ models disturbances acting on the plant, \emph{tracking} is obtained when $\y$ represents the ``error'' between a given plant's output and a reference trajectory dependent of $w$, and some  \emph{robust control} problems are obtained whenever $w$ represents uncertain parameters or unmodeled dynamics. As customary in the output regulation literature,   we assume here that the exogenous signal $w$ belongs to the set of solutions of an \emph{exosystem} of the form
\begin{equation}
\label{s:exo}
\dot{w}= s(w) ,
\end{equation}
originating in a compact invariant subset $\rW$ of $\R^\dw$.

Output regulation is subject to the following taxonomy. \emph{Asymptotic regulation}   denotes the case in which the control objective is to ensure $\lim_{t\to\infty} y(t)= 0$. \emph{Approximate regulation}   denotes the case in which the control objective is relaxed to $\limsup_{t\to\infty}|\y(t)|\le \vep\sr$, with $\vep\sr$ that represents some performance specification or optimality condition. \emph{Practical regulation} refers to the case in which $\limsup_{t\to\infty}|y(t)|$ can be reduced arbitrarily by opportunely tuning the regulator. When one of the above control objectives is achieved in spite of uncertainties in the plant's model, we call it \emph{robust regulation}. When some learning mechanism is introduced to compensate for uncertainties in the exosystem, the problem is typically referred to as \emph{adaptive regulation}. 
%
Asymptotic output regulation is a rich research area with a well-established theoretical foundation. For linear systems  a complete formalization and solution of the problem has been given in the mid 70s  in the seminal works by Francis, Wonham and Davison (see e.g. \cite{Francis1976,Davison1976}), where the well-known \emph{internal model principle} was first stated. Asymptotic output regulation for (single-input-single-output) nonlinear systems has been under investigation since the early 90s, first in a local context \cite{Huang1990,Isidori1990,Huang1995,Byrnes1997}, and lately in a purely nonlinear framework \cite{Byrnes2003,Byrnes2004,Marconi2007} based on the  ``non-equilibrium'' theory \cite{Byrnes2003b}. In more recent times, asymptotic regulators have been also extended to some classes of multivariable nonlinear systems (see e.g. \cite{Wang2016,WanIsiLiuSu,Bin2019}).

One of the major limitations of the existing  asymptotic regulators is their complexity: the sufficient conditions under which asymptotic regulation is ensured are typically expressed by equations whose analytic solution becomes a hard (if not impossible) task even for ``simple'' problems, with the consequence that the construction of the regulation quickly becomes  unfeasible. As conjectured in \cite{Bin2018robustness}, moreover, even if a regulator can be constructed, asymptotic regulation remains a fragile property that is lost at front of the slightest plant's or exosystem's perturbation. This, in turn, motivates the interest towards approximate, practical and adaptive solutions, sacrificing   asymptotic convergence   to gain robustness and practical feasibility. Among the approaches to approximate regulation it is worth mentioning \cite{Marconi2008,Astolfi2015}, whereas practical regulators can be found in \cite{Isidori2012,Freidovich2008,Bin2019}. Adaptive designs of regulators can be found, e.g., in \cite{Serrani2001,DelliPriscoli2006,Pyrkin2018}, where linearly parametrized internal models are constructed in the context of adaptive control, in \cite{Bin2018d} where discrete-time adaptation algorithms are used in the context of multivariable linear systems, and in \cite{Forte2017,Bin2018,BinBerMar} where adaptation of a nonlinear internal model is approached as a system identification problem.

A further limitation, present in most of the aforementioned designs and representing a major obstacle  to practical implementation, is that the stabilization techniques used in the regulator employ control ``gains'' that need to be taken very large to ensure closed-loop stability,  resulting in undesired ``peaking'' phenomena in the transitory, amplification of noise, and exaggerate strength and rigidity in the counteraction of disturbances. Moreover, the introduction of internal model units and adaptation mechanisms   typically leads to a further increase of the gain, namely one has to ``pay'' in terms of stabilization for introducing  additional complexity potentially leading to better asymptotic performance. This, in turn, makes  more naive controllers preferable despite the lower asymptotic performance. 
In the practical approach of \cite{Freidovich2008}, initially developed to robustify ideal feedback-linearization designs, the stabilizing action does not necessarily employ high gains, and the \emph{high-gain} part is shifted to an additional \emph{extended} observer, with the result that the typical problems linked to high-gain control mentioned above are mitigated. Extended observers have also been extensively studied in the context of disturbances attenuation, see for instance \cite{Li2014}. The approach of \cite{Freidovich2008}, originally dealing with practical stabilization, has been   extended in \cite{WanIsiLiuSu} to a class of multivariable systems, where the controller is augmented by an internal model which also allows one to deal with (possibly asymptotic) output regulation problems. Although theoretically appealing, the design of \cite{WanIsiLiuSu} is not constructive, in the sense that only an existence result of the internal model unit is given and no constructive design conditions   are given even for simple problems. 

In this paper we start from the idea of \cite{WanIsiLiuSu} and \cite{Freidovich2008} to construct a regulator for multivariable nonlinear systems embedding an internal model unit that is adapted at run time on the basis of the measured closed-loop signals. Compared to \cite{Freidovich2008}, we consider multivariable regulation problems  rather than single-variable practical stabilization. Compared to \cite{WanIsiLiuSu}, we confer on the internal model unit the ability to adapt online, thus proposing a control solution which is constructive and does not rely on fragile analytical conditions as typically required by non-adaptive designs. Besides, unlike in \cite{WanIsiLiuSu}, we ensure that the parameters of the controller are fixed a priori independently from the added internal model. On the heels of \cite{Bin2018d,Forte2017,Bin2018}, and contrary to canonical adaptive control designs, adaptation is not carried by means of ``ad hoc'' algorithms developed under structural assumption on the internal model unit and by means of Lyapunov-like arguments; rather we approach  the adaptation of the internal model as a \emph{system identification} problem, where the \emph{best} model matching with the measured data and performance needs to be identified. We thus allow for different identification schemes to be used, by individuating a set of sufficient stability conditions that they need to satisfy to be used within the framework. As in \cite{Bin2018d}, we consider here identifiers that are \emph{discrete-time}, which turn the closed-loop system into a \emph{hybrid system}. Despite the additional complexity in the analysis, this choice is motivated by the fact that identification schemes are typically discrete-time, and that in this way we also structurally support adaptive mechanisms working on sampled data. 

 The paper is organized as follows. In Section \ref{sec:framework} we describe the standing assumptions and we further discuss the previous results and the contribution of the paper. In Section \ref{sec:regulator} we present the proposed regulator and in Section \ref{sec:MainResult} we state the main result of the paper, proved later in Section \ref{sec:pf_thm}. In Section \ref{sec:DesignOfIdentifiers} we construct some identifiers for linear and nonlinear parametrizations and, finally, in Section \ref{sec:Example} we present a numerical example.

\emph{Notation:} We denote by $\R$ and $\N$ the sets of real and natural numbers, $\Rplus:=[0,\infty)$ and $\Nz:=\N\setminus\{0\}$. When the underlying metric space is clear, we denote by $\ball_\varrho$ the open ball of radius $\varrho$ and, if $\cB$ is a set, we denote by $\ball_\varrho^\cB$ the open ball of radius $\varrho$ around $\cB$. If $S$ is a set, $\closed S$ denotes its closure. If $B$ is another set, $S\subseteq B$ (resp. $S\subset B$) means $S$ is contained (resp. strictly contained) in $B$. 
Norms are denoted by $|\cdot|$. If $\cA\subset\R^n$, $|x|_\cA := \inf_{a\in\cA}|x-a|$ denotes the usual distance of $x\in\R^n$ to $\cA$. 
For $x:\N\to\R^n$ (resp. $x:\R\to\R^n$), we let $|x|_j:=\esssup_{i\le j}|x(i)|$ (resp. $|x|_t:=\esssup_{s\le t}|x(s)|$). If $\cA\subset\R^n$, we let for convenience $|x|_{\cA,j}:= | |x|_\cA |_j$ (resp. $|x|_{\cA,t}:= | |x|_\cA |_t$).
If $A_1,\dots, A_m$ are matrices, we let $\diag(A_1,\dots,A_m)$ and $\col(A_1,\dots,A_m)$ their block-diagonal and column concatenation respectively. We denote by $\SPD_n$ the set of positive semi-definite symmetric matrices of dimension $n$. A function $\kappa:\Rplus\to\Rplus$ is said to be of class-$\cK$ ($\kappa\in\cK$) if it is continuous, strictly increasing, and $\kappa(0)=0$. A  function $\kappa\in\cK$ is said to be of class-$\cK_\infty$ ($\kappa\in\cK_\infty$) if $\lim_{s\to\infty}\kappa(s)=\infty$. A function $\beta:\Rplus\x\Rplus\to\Rplus$ is said to be of class-$\cK\cL$ if $\beta(\cdot,t)\in\cK$ for each $t\in\Rplus$ and, for each $s\in\Rplus$, $\beta(s,\cdot)$ is continuous and strictly decreasing to zero as $t\to\infty$. By $F:A\setto B$ we denote  set-valued maps. A function $f:\R^n\to\R^m$ is $\cC^k$ if $k$ times continuously differentiable. If $h:\R^n\to\R$ is $\cC^1$ and $f:\R^n\to\R$, for each $i\in\{1,\dots,n\}$ we denote by $L_{f}^{(x_i)} h$ the map $x\mapsto L_{f(x)}^{(x_i)} h(x):=\partial h/\partial x_i(x) \cdot f(x)$. When the superscript $(x_i)$ is obvious, it is omitted.

In this paper we deal with \emph{hybrid systems}, i.e., systems that combine discrete- and continuous-time dynamics. They are formally described by equations of the form \cite{Goebel2012book}
\begin{equation}\label{s:hs0}
\Sigma: \left\{ \begin{array}{lclrl}
\dot x &= & F(x,u) && (x,u)\in  C \\
x^+ &= & G(x,u) && (x,u)\in D
\end{array}\right.
\end{equation}
where $F$ and $G$ denote the {\em flow} and {\em jump} maps and $C$ and $D$ the sets in which flows and jumps are allowed. 
Solutions to \eqref{s:hs0} are defined over \emph{hybrid time domains}.
A \emph{compact hybrid time domain} is a subset of $\Rplus\x\N$ of the form $\cT=\cup_{j=0}^{J-1} [t_j,t_{j+1}]\x\{j\}$ for some finite $J\in\N$ and $0=t_0\le t_1\le\dots\le t_J\in\Rplus$. A set $\cT\subseteq\Rplus\x\N$ is called a \emph{hybrid time domain} if for each $(T,J)\in\Rplus\x\N$ $\cT\cap[0,T]\x\{1,\dots,J\}$ is a compact hybrid time domain. 
If $(t,j),(s,i)\in \cT$, we write $(t,j)\preceq (s,i)$ if $t+j\le s+i$. For any $(t,j)\in\cT$, we let   $t^j=\sup_{t\in\R}(t,j)\in\cT$, $t_j:=\inf_{t\in\R}(t,j)\in\cT$ and $j_t$ and $j^t$ in similar way. 
%
%
A function $x:\cT\to\cX$ defined on a hybrid time domain $\cT$ is called a \textit{hybrid arc} if $x(\cdot,j)$ is locally absolutely continuous for each $j$. A \emph{hybrid input} is a hybrid arc that is locally essentially bounded and Lebesgue measurable. A \emph{solution pair} to \eqref{s:hs0} is a pair $(x,u)$, with $x$ a hybrid arc and $u$ a hybrid input, that satisfies such equations.  We call a solution pair   {\em complete} if its time domain is unbounded.  We let $\dom x$ denote the domain of $x$, and $\jumps x\subseteq\N$ the set of $j$ such that $(t,j)\in\dom x$ for some $t\in\R$. In order to simplify the notation, we omit the jump (resp. flow) equation when the considered system has only continuous-time (resp. discrete-time) dynamics. If $x$ is constant during flows, we neglect the ``$t$'' argument and we write $x(j)$, which we identify with the map $j\mapsto x(t_j,j)$. In the same way we write $x(t)$ for hybrid arcs that are constant during jumps, and we identify $x$ with the map $t\mapsto x(t,j_t)$.
For a hybrid input $u:\dom u\to\cU$,  $\Gamma(u):=\{(t,j)\in \dom u\st (t,j+1)\in \dom u\}$, and for   $(t,j)\in\dom u$ we let $|u|_{(t,j)}:=\max\{   \sup_{(s,i)\in\Gamma(\dom u),\,(0,0)\preceq(s,i)\preceq(t,j)} |u(s,i)|, \middlebreak \esssup_{(s,i)\in\dom u\setminus\Gamma(\dom u),(0,0)\preceq (s,i)\preceq (t,j)} |u(s,i)|\}$. We also let $|u|_{A,(t,j)}:=\big||u|_A \big|_{(t,j)}$ and $|u|_\infty:=\limsup_{t+j\to\infty} |u|_{(t,j)}$. In the paper, ``ISS'' stands for ``input-to-state stability''. 

\section{The Framework}\label{sec:framework}

\subsection{Standing Assumptions}\label{sec:framework:assumptions}
We consider the problem of adaptive output regulation for systems of the form \eqref{s:plant}, \eqref{s:exo} under a set of assumptions detailed hereafter. 
\begin{ass}\label{ass:regularity}
The function $f$ is locally Lipschitz and the functions $q$ and $b$ are $\cC^1$ with locally Lipschitz derivative.\qedenv
\end{ass}
\begin{ass}\label{ass:minPhase}
	There exists a $\cC^1$ map $\pi:\R^\dw\to\R^\dz$ satisfying
	\begin{equation*}
	L_{s(w)} \pi(w) = f(w,\pi(w),0)
	\end{equation*}
	in an open set including $\rW$,  such that the system
	\begin{equation} 
	\label{s:dyn_wz} 
	\dot{w} =  s(w),\qquad
	\dot{z} =  f(w,z,x) 
	\end{equation}
	is ISS relative to the compact set 
	$
	\cA = \{(w,z)\in \rW\times \R^\dz \: : \: z= \pi(w) \}
	$
	with respect to the input $x$ with locally Lipschitz asymptotic gain. More precisely, there exist $\beta_0\in\cK\cL$ and a locally Lipschitz $\rho_0\in\cK$ such that every solution pair to \eqref{s:dyn_wz} originating in $\rW\x\R^\dz$  satisfies
	\begin{equation*}
	|(w(t),z(t))|_\cA \le \max\left\{ \beta_0(|(w(0),z(0)|_\cA,t),\, \rho_0(|x|_t) \right\},
	\end{equation*}
	for every $t\in\Rplus$.\qedenv
\end{ass}
\begin{ass}\label{ass:b}
There exist a known  nonsingular matrix $\bb\in\R^{\dy\x\dy}$ and a known scalar $\mu \in(0,1)$ such that the following holds
\begin{equation}\label{e:bb}
|(b(w,z,x)-\bb)\bb\inv|\le 1-\mu 
\end{equation}
for all $(w,z,x)\in\rW\x\R^\dz\x\R^\dx$.\qedenv
\end{ass}

Assumption A\ref{ass:minPhase} is a minimum-phase assumption, asking that the zero dynamics of \eqref{s:plant}, \eqref{s:exo}, described by
\begin{equation}\label{s:zerodyn} 
\dot w  =  s(w),\quad 
\dot z  =  f(w,z,0), 
\end{equation}
has a steady state of the form $z=\pi(w)$ that is compatible with the control objective $\y=0$ and that is \emph{robustly} asymptotically stable. Minimum-phase is a customary (although not necessary) assumption in the literature and, in this respect, A\ref{ass:minPhase} represents a strong  minimum-phase assumption, where the adjective ``strong'' refers to the ISS requirement. Nevertheless, we remark that, by means of well-known arguments (see e.g. \cite{Marconi2007,Byrnes2003b}), A\ref{ass:minPhase} could be relaxed to assume that $\cA$ is ``just'' locally exponentially stable for \eqref{s:zerodyn}, provided that the only component of $x$ that affects $f$ is $\y=Cx$. 
A\ref{ass:b}  is instead a stabilizability assumption taken from \cite{WanIsiLiuSu,Freidovich2008} and asking the designer to have available an estimate $\bb$ of $b(w,z,x)$ which captures enough information on its behavior. A\ref{ass:b}, in particular, implies that $b(w,z,x)$ is nonsingular at each $(w,z,x)$. We also remark that A\ref{ass:b} could be weakened to a ``local'' version, i.e. requiring that a  pair $(\bb,\mu)$ fulfilling \eqref{e:bb} exists for each compact subset of $\rW\x\R^\dz\x\R^\dx$.

\subsection{Previous Approaches}\label{sec.framework.previous}
%
There follows by the structure of \eqref{s:plant}, \eqref{s:exo} that, under A\ref{ass:minPhase}, the problem of asymptotic regulation  could be in principle solved    by a control law of the kind
\begin{equation}\label{d:u_ideal}
u = -b(w,z,x)\inv q(w,z,x) + b(w,z,x)\inv k(x), 
\end{equation}
where the term $-b(w,z,x)\inv q(w,z,x)$ represents a non-vanishing ``feedforward'' action compensating for the influence of the dynamics of $(w,z)$ on $\dot x$, and  $k(x)$ is a stabilizing control action vanishing with $x$. However, \eqref{d:u_ideal} cannot be directly implemented even if the whole state $(z,x)$ were accessible, as it anyway would require  $w$ to be measured and the functions $q$ and $b$ to be perfectly known.
To overcome those issues, in \cite{Freidovich2008}  the authors proposed a dynamic regulator in a single-variable context (i.e. $\dy=1$) where $b$ and $q$ in \eqref{d:u_ideal} are approximated by functions $x\mapsto\hat q(x)$ and $x\mapsto\hat b(x)$ of $x$ only,  and an \emph{extended observer} is introduced to provide an estimate $\hat x$ of $x$ and to compensate for the  mismatch of $\hat b$ and $\hat q$ with the actual quantities. The control action was taken as
\begin{equation}\label{d:u_khalil}
u := \sat\left( \hat b(\hat x)\inv \big(-\hat q(\hat x) + k(\hat x) - \hat\sigma\big) \right),
\end{equation}
where $\sat$ is a suitably chosen saturation function and $\hat\sigma$ is the term of the extended observer compensating for the mismatch between $(\hat q(\hat x),\hat b(\hat x))$ and $(q(w,z,x),b(w,z,x))$.
This regulator was proved to recover the performance of the ideal control law \eqref{d:u_ideal}   theoretically  as closely and quickly as desired, by increasing the observer gains accordingly. Nevertheless, the regulator of \cite{Freidovich2008} does not embed any process which is  able to generate the ideal feedforward term $-b(w,z,x)\inv q(w,z,x)$, which indeed can be only approximated by the extended observer. Therefore, the attained regulation result is only practical, with the observer gains that must be taken high enough to accommodate the desired asymptotic bound. This design thus has two main drawbacks. First, the ideal steady state in which $\y=0$ is not a trajectory of the system and, as such, it is not  stable, so that a considerable transitory is possible even if the system is initialized close to the desired operating point. Second, good performance are only obtained by increasing the observer gains accordingly. As the observer gains grow, however,   the peaking and the noise amplification grow, so that a compromise between regulation performance and high gain must be sought. A remarkable property of this approach is   that the stabilizing action $k(x)$ is not forced to be ``high-gain'' and is fixed a priori in the ``ideal'' controller \eqref{d:u_ideal}.

On the other side, when $\dx=\dy=1$, it was shown in \cite{Marconi2007}  that, under A\ref{ass:minPhase} and if $b(w,z,x)$ is lower bounded by a positive constant,  the problem of asymptotic output regulation for  \eqref{s:plant}, \eqref{s:exo} can  always be  solved by means of  a controller of the form 
\begin{equation}\label{s:ctr_sicon}
\begin{array}{lcl}
\dot{\eta} &=& F\eta + Gu  \\
u&=&\gamma(\eta) + \kappa(x),
\end{array}
\end{equation}
with state $\eta\in\R^\deta$, $\deta=2(\dz+\dw+1)$,  $(F,G)$ a controllable pair with $F$ a Hurwitz matrix,  and with $\gamma:\R^\deta \to \R$ and $\kappa:\R \to \R$ suitably defined continuous functions. The term $\kappa(x)$ plays here the same role as $k(x)$ in \eqref{d:u_ideal}, while the term $\gamma(\eta)$ is meant to reproduce the feedforward action $-b(w,z,x)\inv q(w,z,x)$ at the steady state. For this reason, the restriction of \eqref{s:ctr_sicon} to the set in which $x=0$, namely
\begin{equation*}
\dot\eta = F\eta + G\gamma(\eta),\qquad u=\gamma(\eta),
\end{equation*}
is called the \emph{internal model} unit, as it is able to generate  the ideal feedforward action making the set where $\y=0$ invariant (property that the regulator of \cite{Freidovich2008} does not have). This approach, however, has two main drawbacks: the stabilizing action $k(x)$ is necessarily high-gain to bring the system  close to the steady state where $\gamma(\eta)$ behaves as desired, and even if $\gamma$   always   exists, no analytical or numerical method exists to construct it even for simple problems.

In   \cite{WanIsiLiuSu}, the authors extended both the approaches of \cite{Freidovich2008,Marconi2007} described above to the class of systems \eqref{s:plant}, \eqref{s:exo}. The approach of \cite{WanIsiLiuSu}, in particular, is based on an extension of the  extended observer of \cite{Freidovich2008} to multivariable systems, where   $\hat b$ is taken constant in \eqref{d:u_khalil} and equal to $\bb$ of A\ref{ass:b}, and the term $\hat b(\hat x)\inv \hat q(\hat x)$ is substituted by the output $\gamma(\eta)$ of an internal model unit of the kind \eqref{s:ctr_sicon}, appropriately extended to fit the multivariable setting. Then, $u$ is   taken as
\begin{equation}\label{d:Wang_u}
u=\gamma(\eta) + \bb \inv \big(  -\sat(\hat{\sigma}) +k(x)  \big) .
\end{equation}
 Compared to \cite{Freidovich2008}, this design is potentially asymptotic (whenever \eqref{s:ctr_sicon} is chosen correctly), thus possibly achieving $\y\to 0$ without taking the observer gains inconveniently large. Compared to \cite{Marconi2007}, apart from the  extension to multivariable normal forms, the approach of \cite{WanIsiLiuSu} allows one to use stabilization control actions that are not high-gain. 
However, the problems related to the construction of $\gamma$ inherited from \cite{Marconi2007} persist, with the consequence that, although theoretically appealing, the approach of \cite{WanIsiLiuSu} is not constructive. Besides, the saturation level of the map $\sat$ depends on the choice of internal model, and in particular of $\gamma$ itself, and on the initial error in the initialization of $\eta$ relative to its (unknown) ideal steady state.
 Some existing  methods to approximate   $\gamma$  have been proposed in \cite{Marconi2008}, yet their implementation remains tedious and the computational complexity   easily grows with the desired precision and the dimension of the problem.
 Otherwise,  adaptive designs exist  that tune $\gamma$ online (see \cite{Pyrkin2018,BinBerMar}), yet they are far from a definite answer and are all based on high-gain stabilization.

\subsection{Contribution of the paper}
In this paper we present a regulator embedding an adaptive internal model unit and non-high-gain stabilization actions, by thus merging all the desired properties mentioned before.  Adaptation is cast as a discrete-time system identification problem \cite{Ljung1999} defined over samples of the closed-loop system trajectories. Instead of developing a single ad hoc adaptation algorithm, we give  sufficient conditions under which arbitrary identification schemes can be used. We then specifically develop the relevant case of weighted \emph{least squares} for linear parametrizations and \emph{mini-batch algorithms} for nonlinear parametrizations, thus embracing many existing and frequently-used techniques performing white- and black-box identification. The proposed regulator is proved to achieve both practical and approximate regulation, with an asymptotic bound that is directly related to the \emph{prediction capabilities} of the identifier. Hence, the result becomes asymptotic whenever the identified model is perfect.   Compared to \cite{Freidovich2008}, the proposed regulator has the ability to learn and employ an internal model unit reproducing the ideal feedforward action making the set in which $\y=0$ asymptotically stable. Compared to \cite{Marconi2007}, the proposed approach does not rely on high-gain stabilization and, compared to \cite{Marconi2007} and \cite{WanIsiLiuSu}, we introduce adaptation of the internal model, which provides a constructive method to compute $\gamma$ online. Besides, unlike in \cite{WanIsiLiuSu}, the parameters of the controller are fixed a priori based on the plant and exosystem dynamics, and independently from the added internal model, identification and observer units.
%


\section{The Regulator}\label{sec:regulator}
The proposed regulator is a hybrid system described by 
\begin{equation}\label{s:ctr}
\begin{aligned}
&\left\{\begin{array}{lcl}
\dot\tim &=& 1\\
\dot\eta &=& F\eta + G u  \\
\dot{\hat{x}} &=&  A \hat{x} + B(\hat{\sigma} + \bb  u) + \Lambda(\ell) H (y-\hat{x}_1)\\
\dot{\hat{\sigma}} &=& -\bb  \psi(\theta,\eta,u)+\ell^{r+1} H_{r+1} (y-\hat{x}_1)\\
\dot \xi &=& 0 \\
\dot\theta &=& 0
\end{array}\right.\\
&\qquad (\tim,\eta,\hat x,\hat\sigma,\xi,\theta,\y) \in \rC_\tim \x \R^{\deta+\dx+\dy}\x\Xi\x\Theta\x\R^\dy \\[.2em]
&\left\{\begin{array}{lcl}
\tim^+ &=& 0,\quad 
\eta^+  =  \eta  \\
\hat{x}^+ &=&  \hat x,\quad 
\hat{\sigma}^+  =  \hat\sigma \\
\xi^+ &=& \vhi(\xi,\eta,u) \\
\theta^+ &= & \Thmap (\xi) 
\end{array}\right.\\  
&\qquad (\tim,\eta,\hat x,\hat\sigma,\xi,\theta,\y) \in \rD_\tim \x \R^{\deta+\dx+\dy}\x\Xi\x\Theta \x\R^\dy
\end{aligned}  
\end{equation}
and with output
\begin{equation}\label{s:ctr_u}
u= \bb\inv\sat\big(   -\hat\sigma + \kappa(\hat x) \big),
\end{equation}
in which $\bb$, $A$, and $B$ are the same matrices of A\ref{ass:b} and \eqref{s:plant}, \eqref{s:exo} respectively, $\deta\in \N$, $\Xi$ and  $\Theta$ are finite-dimensional normed vector spaces,  $(F,G)\in\R^{\deta\x \deta}\x\R^{\deta\x\dy}$ and $(\Lambda(\ell),H,H_{r+1})\in\R^{\dx\x\dx}\x\R^{\dx\x\dy}\x\R^{\dy\x\dy}$ are matrices to be defined, $\ell\in\Rplus$ is a control parameter, $\psi:\Theta\x\R^\deta\x\R^\dy\to\R^\dy$, $\vhi:\Xi\x\R^\deta\x\R^\dy\to\Xi$, $\sat:\R^\dy\to\R^\dy$, $\kappa:\R^\dx\to\R^\dy$, $\Thmap:\Xi\to \Theta$ are functions to be designed and, with $\upT,\,\unT\in\Rplus$ satisfying $0<\unT\le\upT$  
\begin{align}\label{d.sets_C_D}
\rC_\tim &:= [0,\upT],& \rD_\tim&:=[\unT,\upT].
\end{align}

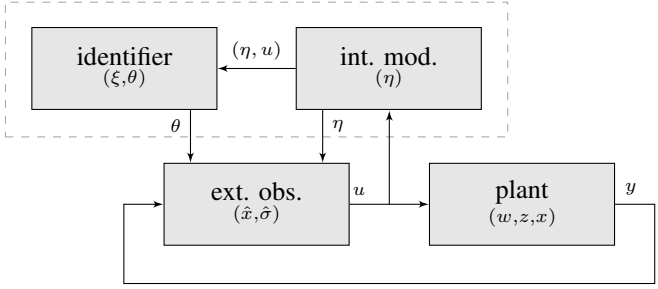
\begin{figure}[t]
	\centering
	\tikzstyle{sys} = [draw,fill=black!10!white,inner sep=2,minimum width=7em,minimum height=3.1em]
	\tikzstyle{clk} = [circle,draw,inner sep=2]
	\tikzstyle{line} = [draw, -latex']
	\begin{tikzpicture}
	\node[sys] (plant) at (0,0) {$\underset{(w,z,x)}{\text{plant}}$};
	
	\node[sys] (stab) [left=3em of plant.west] {$\underset{(\hat x,\hat\sigma )}{\text{ext. obs.}}$};
	\node[sys] (intm) [above=2em of stab.north,xshift=5em] {$\underset{(\eta)}{\text{int. mod.}}$};
	\node[sys] (id) [above=2em of stab.north,xshift=-5em] {$\underset{(\xi,\theta)}{\text{identifier}}$};
	
	\node[draw=black!35!white,dashed,minimum width=19em,minimum height=5.1em,anchor=center,below=0em of intm,yshift=4.1em,xshift=-5em] {};

	\path[line] (plant.east)  -| ($(plant.east)+(1.5em,-3em)$) node[above,pos=.2] {\scriptsize $y$}  -| ($(stab.west)-(1.5em,0)$) -- (stab.west)  ;
	
	\path[line] (stab.east) -- (plant.west) node[above,pos=.12] {\scriptsize $u$};
	
	\path[line] (intm.west)--(id.east) node[above,midway] {\scriptsize $(\eta,u)$};
	
	\path[line] ($(stab.east)+(1.5em,0)$) -- ($(stab.east)+(1.5em,3.55em)$);

	\path[line]   ($(id.south)+(2.5em,0em)$) -- ($(stab.north)-(2.5em,0)$) node[pos=.3,left] {\scriptsize $\theta$};
	\path[line]   ($(intm.south)-(2.5em,0em)$) -- ($(stab.north)+(2.5em,0)$) node[pos=.3,right] {\scriptsize $\eta$};
	
	\end{tikzpicture}
	\caption{Block-diagram of the regulator.}
	\label{fig:ctrl}\vspace{-1em}
\end{figure}
%
%
%
%
%
%
%
%

The regulator, whose block-diagram is depicted in Figure~\ref{fig:ctrl}, is composed of:
a) a purely continuous-time subsystem $(\eta,\hat x,\hat\sigma)$, whose dynamics depends on a parameter $\theta$ that is constant during flows; b) a purely discrete-time subsystem $(\xi,\theta)$ updated at  jump times; c) a hybrid \textit{clock} $\tim$ whose tick  triggers  the updates of the parameter $\theta$.
The definition of the {flow} and {jump} sets $\rC_\tim$ and $\rD_\tim$   allows the usage of any arbitrary, and possibly aperiodic, clock strategy   in which the distance of two successive jumps is lower bounded by $\unT$ and upper bounded by $\upT$. 
The subsystem $\eta$, taking values in $\R^\deta$, plays the role of an \emph{internal model unit}, and is taken of the same form as \eqref{s:ctr_sicon}. The subsystem $(\hat x,\hat \sigma)$, taking values in $\R^{\dx+\dy}$, is an \emph{extended observer} similar to that of \cite{Khalil1998}, but with an additional ``consistency term'' $-\bb\psi(\theta,\eta,u)$ which, as better clarified later, represents the output of the internal model unit. The subsystem $\xi$, taking values in $\Xi$, is the \emph{identifier}, whose updates take place at jump times. The variable $\theta$, taking values in $\Theta$, is the identifier's output, and it is included as a state in \eqref{s:ctr} to formalize the fact that it only changes at jump times. 
In the rest of the section we detail the construction of all these subsystems, along with all the degrees of freedom introduced in \eqref{s:ctr}. In doing so, we make reference to a given arbitrary set of initial conditions for \eqref{s:plant}, \eqref{s:exo} of the form $\rW\x\rZ_0\x\rX_0\subset\R^\dw\x\R^\dz\x\R^\dx$.

\begin{remark}
We underline that, contrary to \cite{Marconi2007,WanIsiLiuSu}, the output $\eta$ of the internal model unit does not enter directly in the definition of $u$ (compare \eqref{s:ctr_u} with \eqref{s:ctr_sicon}, \eqref{d:Wang_u}), but only in the dynamics of $\hat\sigma$ through the map $\psi$. As it will be clarified in the next subsection, unlike  \cite{WanIsiLiuSu},  this allows us to fix the saturation level of \eqref{s:ctr_u} \emph{independently} from the extended observer, the internal model and the identifier.\qedenv
\end{remark}

\subsection{The Clock Subsystem}\label{sec:clock}
The clock dynamics is described by the following equations
\begin{equation}\label{s.clock}
\begin{array}{lclcr}
\dot \tim &=& 1 && \tim\in \rC_\tim\\
\tim^+ &=& 0 && \tim\in\rD_\tim
\end{array}
\end{equation}
in which the sets $\rC_\tim$ and $\rD_\tim$ are defined in \eqref{d.sets_C_D}. By construction, each solution to \eqref{s.clock} has infinite flow intervals and jump times, and each two successive jump times are separated by at least $\unT$ and at most $\upT$ seconds.
Furthermore, by definition of the flow and jump sets of \eqref{s:ctr}, and since the plant \eqref{s:plant} is a purely continuous-time system,  the flow and jump times of the solutions to the resulting closed-loop system  \eqref{s:plant}, \eqref{s:ctr}   are the same as the clock subsystem. 

We stress, moreover, that the equations \eqref{s.clock} do not correspond to the implementation of a single clock strategy. Rather, they model an uncountable family of possible strategies that the designer can implement in the proposed framework to trigger the updates of the discrete-time dynamics of \eqref{s:ctr}. By way of example, a periodic clock strategy with period $T\in[\unT,\upT]$ is a solution to \eqref{s.clock} and, thus, it is a suitable clock strategy. More in general, \emph{every} clock strategy which can be described by the dynamic equations \eqref{s.clock} can be used in the proposed framework. Developing the analysis on \eqref{s.clock}, in turn, permits us to capture all these possible clock strategies at once, without needing to know which one in particular will be implemented. The constants $\unT$ and $\upT$, which are the only degrees of freedom characterizing the clock subsystem, are \emph{arbitrary}. However,  as specified in the forthcoming Theorem~\ref{thm:main},   the rest of the regulator   depends on their value. Moreover, we also underline that a given clock strategy strongly affects the data set that will be made available to the identifier, thus potentially affecting its performance. In this respect, there is no a ``best way'' to choose the clock strategy, which is left here as a degree of freedom to the designer. 

\subsection{The Stabilizing Action}\label{sec:stabilizer}
In this section we fix the functions $\kappa$ and $\sat$ in \eqref{s:ctr_u}. The function $\kappa$ is chosen as any $\cC^1$ function such that the system
	\begin{equation}\label{s:x_kappa}
	\dot x = Ax + B\kappa(x) + B\delta 
	\end{equation}
	is ISS relative to the origin and with respect to $\delta$ with locally linear asymptotic gains. Namely, such that there exist $\beta_x\in\cK\cL$ and a locally Lipschitz  $\rho_x\in\cK$ for which \eqref{s:x_kappa} satisfies
	\begin{equation*}
	|x(t)|\le \max\{ \beta_x(|x(0)|,t),\,\rho_x(|\delta|_t)\}
	\end{equation*}
	for all $t\in\Rplus$.  For instance, $\kappa$ can be chosen as $\kappa(x)=Kx$, with  $K\in \R^{\dy\x\dx}$ such that $A+BK$ is Hurwitz.
There follows from A\ref{ass:minPhase}  that the system \eqref{s:dyn_wz}, \eqref{s:x_kappa} is ISS relative to the set
\begin{equation*} 
\cB:=\cA\x\{0\}=\big\{   (w,z,x)\in\cA\x\R^\dx\st  x=0 \big\} 
\end{equation*}
and with respect to the input $\delta$. 
Let $(\rZ_0,\rX_0)$ be the sets of initial conditions for \eqref{s:plant} and $\varrho_0>0$ such that 
$
 (\rW\x\rZ_0\x\rX_0) \subseteq \closed\ball_{\varrho_0}^\cB .
$
With $\bar{\delta}$ and $\varrho_1>\varrho_0$ arbitrary positive scalars, there exists a compact set $\Omega_0\subset\R^\dw\x\R^\dz\x\R^\dx$ satisfying
\begin{equation}\label{e:Omega_0}
(\rW\x\rZ_0\x\rX_0)\subseteq   \closed\ball_{\varrho_0}^\cB\subset \closed\ball_{\varrho_1}^\cB \subset \Omega_0,
\end{equation}
and such that every trajectory of the system \eqref{s:dyn_wz}, \eqref{s:x_kappa} originating in $\closed\ball_{\varrho_1}^\cB$ and with an input $\delta$ satisfying $|\delta|_\infty\le\bar\delta$ is complete, and fulfills $(w(t),z(t),x(t))\in\Omega_0$ for all $t\in\Rplus$.

Let $c:\rW\x\R^\dz\x\R^{\dx}\to\R^\dy$ be defined as
\begin{equation*}
c(w,z,x) := - b(w,z,x)\inv q(w,z,x),
\end{equation*}
and, with $\varrho_2>0$ arbitrary, let $\satlev$ be any constant fulfilling
\begin{equation}\label{d:satlev}
\satlev  \ge \max_{(w,z,x)\in\Omega_0} \big| \bb c(w,z,x) + \bb b(w,z,x)\inv\kappa(x) \big|+\varrho_2.
\end{equation}
Then we define   $\sat(\cdot)$ as any $\cC^1$ bounded function satisfying\footnote{All the subsequent results can be proved even if $\sat$ is differentiable a.e.; the $\cC^1$ requirement, in turn, is asked to simplify the forthcoming analysis.}
\begin{equation}\label{d:sat}
\begin{aligned}
&0\le |\sat'(s)| \le 1 & \forall &s\in\R^\dy \\
&\sat(s) = s & \forall & s\in \closed\ball_{\satlev}.
\end{aligned}
\end{equation}
\begin{remark}\label{rmk:sat}
The definition of $\satlev$ requires the knowledge of a bound on the maximum value that the functions $\bb c(w,z,x)$ and $\bb b(w,z,x)\inv\kappa(x)$ attain  in $\Omega_0$. While knowing a bound of $c(w,z,x)$ is  a \emph{quantitative} information related to the plant, and in particular on the ideal feedforward control action in a neighborhood of set $\cB$, the knowledge of a bound for $\bb b(w,z,x)\inv \kappa(x)$ does not ask for any additional information. In fact, $\kappa$ is known to the designer, while we have $|\bb b(w,z,x)\inv| \le \mu\inv$ for all $(w,z,x)\in\rW\x\R^\dz\x\R^\dx$ with $\mu$ defined  in A\ref{ass:b}. Indeed, $\bb b(w,z,x)\inv = ( I + (b(w,z,x)-\bb)\bb\inv )\inv$, so that by \cite[Proposition 10.3.2]{CampbellPseudoInvBook}, $|\bb b(w,z,x)\inv|\le (1-|(b(w,z,x)-\bb)\bb\inv|)\inv \le \mu\inv$.\qedenv
\end{remark}

\subsection{The Internal Model Unit}\label{sec:internalmodel}
The restriction of $c$ on $\cB$, which we denote by
\begin{equation}\label{d:ustar}
u\sr(w) := c(w,\pi(w),0),
\end{equation}
represents the steady-state value of the ideal feedforward action $c$ when $\y$ vanishes, i.e.,  $u\sr(w)$ is the control action that makes the set $\cB$ invariant for \eqref{s:plant}, \eqref{s:exo}. 
The internal model unit $\eta$  is a system constructed  to generate $u\sr(w)$ when $\y=0$, and its construction follows the approach \eqref{s:ctr_sicon} of \cite{Marconi2007}, where the dimension $\deta$ of the state $\eta$ is chosen as $\deta = 2(\dw+\dz+1)$, and the pair $(F,G)$ is taken as a real realization  of any complex pair $(F_c,G_c)$ of dimension $\dw+\dz+1$, with $G_c$ a matrix with non zero entries and $F_c$ a  matrix whose eigenvalues  have sufficiently negative real part. More precisely, this choice is legitimated by the following lemma, which is a direct consequence of   \cite{Marconi2007}.
\begin{lemma} 
	\label{lem:SICON}  
	Suppose that A\ref{ass:minPhase} holds and let $\deta = 2(\dw+\dz+1)$. Then  there exist a controllable pair $(F,G)\in \R^{\deta\times \deta}\x\R^{\deta\times \dy}$, with $F$ a Hurwitz matrix, and continuous maps $\tau : \R^\dw\to \R^\deta$ and $\gamma:\R^\deta \to \R^\dy$ such that
	\begin{equation}
	\label{d:gamma}
	\gamma\circ \tau(w) = u\sr(w) \qquad \forall w \in \rW 
	\end{equation} 
	and, for every  input $(x,\delta_1,\delta_2)\in\R^\dx\x\R^\deta\x\R^\deta$ satisfying $|\delta_1|\le \rho_{\delta_1}|(w,z,x)|_\cB$ for some  $\rho_{\delta_1}>0$, the system 
	\begin{equation}
	\label{s:zerodyn_eta} 
	\begin{array}{lcl}
	\dot{w}&=& s(w) \\ 
	\dot{z}&=& f(w,z,x) \\ 
	\dot{\eta}&=& F \eta +G u\sr(w)  + \delta_1+\delta_2
	\end{array} 
	\end{equation} 
	is forward complete and it is ISS relative to the set
	\begin{equation*} 
	\cD:=  \big\{ (w,z,\eta)\in \cA\x \R^\deta \st  \eta=\tau(w) \big\},
	\end{equation*}
	with respect to the input $(x,\delta_2)$ with locally Lipschitz asymptotic gains.\qedenv 
\end{lemma}

Lemma \ref{lem:SICON} implies the existence of  $\beta_1\in\cK\cL$ and a locally Lipschitz $\rho_1\in\cK$ such that every solution pair to \eqref{s:zerodyn_eta} originating in $\rW\x\R^\dz\x\R^\deta$   satisfies
\begin{equation*}
\begin{aligned}
|(w(t),&z(t),\eta(t))|_\cD \\&\le \max\big\{  \beta_1(|(w(0),z(0),\eta(0))|_\cD,t), \rho_1(|(x,\delta_2)|_t) \big\},
\end{aligned}
\end{equation*}
for all $t\in\Rplus$. 
System \eqref{s:zerodyn_eta} is the zero dynamics, relative to the input-output pair $(u,\y)$, of the plant augmented with the system $\eta$, and the result of Lemma \ref{lem:SICON} states that, in the zero dynamics set $\cD$, we have $\gamma(\eta) = u\sr(w)$,
i.e., the set 
\begin{equation}\label{d:cE}
\cE:=\big\{ (w,z,x,\eta)\in\cB\x\R^\deta \st (w,z,\eta)\in\cD\big\}
\end{equation}
is made invariant for the augmented system with $\delta_2=0$ by the input $u=\gamma(\eta)$. The role of the input $\delta_2$ will be clarified in the forthcoming sections. The map $\gamma$ in \eqref{d:gamma}, which is the same as in \eqref{s:ctr_sicon}, is introduced here to support the subsequent analysis and we stress that \emph{it is not used} in the construction of the regulator.
The actual term through which $\eta$ affects the extended observer is given by the  consistency term $-\bb\psi(\theta,\eta,u)$,   defined later in Section \ref{sec:observer}.

%
%

\subsection{The Identifier}\label{sec:Identifier}
The identifier is a discrete-time system aimed to produce an estimate of the map $\gamma$ introduced in the previous section. The estimation of $\gamma$ is cast here as a system identification problem \cite{Ljung1999}, and the particular design of the degrees of freedom $(\Xi,\vhi,\Theta,\Thmap)$ corresponds to a choice of a given identification algorithm. What is the \emph{right} identification algorithm to use, in turn, is a question whose answer strongly depends on the a priori information that the designer has on the plant, on the exosystem, and on the kind of  uncertainties expected in the different models. In this paper we do not intend to limit to a single choice, which may be good in some settings and inappropriate in others, and we rather   give   a set of sufficient conditions, gathered in what we called the \emph{identifier requirement}, representing the stability and optimality properties that any identification algorithm needs to possess to be used in the framework. We postpone examples of identifiers to Section \ref{sec:DesignOfIdentifiers}.

The identification problem underlying the design of the identifier is cast on the samples of the   following \emph{core process} 
\begin{equation}\label{s:core_process}
 \begin{aligned}
&\left\{\begin{array}{lcl}
\dot\tim &=&1\\
\dot w &=& s(w)\\
\end{array}\right. & (\tim,w)&\in \rC_\tim\x \rW\\ 
&\left\{ \begin{array}{lcl}
\tim^+ &=&0\\
 w^+ &=& w\\
\end{array}\right. & (\tim,w)&\in \rD_\tim\x \rW,
\end{aligned}
\end{equation}
with outputs
\begin{equation}\label{s:core_process_out}
\begin{aligned} 
\win(j)   &:=  \tau(w(t^j,j)), & 
\wout(j)  &:=  u\sr(w(t^j,j)),
\end{aligned}
\end{equation}
where  $u\sr$ and $\tau$ are defined respectively in \eqref{d:ustar} and \eqref{d:gamma}. According to Lemma \ref{lem:SICON}, $\win$ and $\wout$ are linked by 
\begin{equation}\label{e.wout_gamma_win}
\wout = \gamma(\win),
\end{equation} 
and the aim of the identifier is thus to find the  model
$\hat\gamma$  
for which the input-output data pairs $\{(\win(j),\wout(j))\}_{j\in\N}$ fit \emph{at best} (in a way made precise later) the regression \eqref{e.wout_gamma_win}. 

The first step in the construction of the identifier is the definition of a \emph{model set} $\cM$, which is a space of functions where $\hat \gamma$ is supposed to range. As customary in the system identification literature, and due to clear implementation constraints, we limit here to the case in which $\cM$ is finite-dimensional. This, in turn, allows us  to parametrize $\hat{\gamma}$ by a parameter $\theta$ ranging in a finite-dimensional   vector space  $\Theta$, obtaining
\begin{equation}\label{d:model_set}
\cM = \big\{ \hat\gamma(\theta,\cdot) :\R^\deta\to\R^\dy\st \theta\in\Theta \big\}.
\end{equation}
The choice of the model set, and hence of $\Theta$, is guided by the available knowledge on the core process \eqref{s:core_process}-\eqref{s:core_process_out} and, in particular, on the expected relation \eqref{e.wout_gamma_win} between $\win$ and $\wout$, ideally given by the unknown map $\gamma$ (see Lemma \ref{lem:SICON}). Depending on the amount of information available, $\cM$ may range from a very specific set of functions, such as linear regressions, to a space of \emph{universal approximators}, including for instance \emph{Wavelet bases} or \emph{Neural Networks} \cite{Sjoberg1995}.  

Once $\cM$ and $\Theta$ are fixed, a \emph{cost function} is defined on the input-output data set generated by \eqref{s:core_process}-\eqref{s:core_process_out}, so as to assign to each model $\hat\gamma(\theta,\cdot)$ a quantitative value describing how well it fits. In particular, for each solution $(\tim,w)$ to the core process \eqref{s:core_process} and for each $j\in\dom(\tim,w)$ we define the functional
\begin{equation}\label{d:cost_functional}
\cJ_{(\tim,w)}(j,\theta) := \sum_{i=0}^{j-1} \costf\big( \vep(\theta,w(t^i)), i, j\big) + \rho(\theta),
\end{equation}
in which 
\begin{equation}\label{d:vep}
\vep(\theta,w) := u\sr(w) - \hat\gamma(\theta,\tau(w))
\end{equation} 
denotes the \emph{prediction error} attained by the model $\hat\gamma(\theta,\cdot)\in\cM$  along the solution $(\tim,w)$ of \eqref{s:core_process},  $\costf:\R^\dy\x\N^2\to\Rplus$ is a positive function representing the local weight assigned to the term $(\vep(\theta,w(t^i)),i,j)$ in the sum, and $\rho:\Theta\to\Rplus$ is a (possibly zero) \emph{regularization function}. The particular choice of $\costf$ and $\regf$, which is left as a degree of freedom to the designer, characterizes the selection criteria for the best model $\hat\gamma(\theta,\cdot)$.
%
With   \eqref{d:cost_functional} we associate the set-valued map
\begin{equation*}
\optmap_{(\tim,w)}(j) := \argmin{\theta\in\Theta} \cJ_{(\tim,w)}(j,\theta),
\end{equation*}
representing, at each $j$, the set of optimal parameters according to \eqref{d:cost_functional}. In these terms, the identifier  goal  reduces to find, at each $j\in\N$, an \emph{optimal parameter} $\theta\sr(j)\in\optmap_{(\tim,w)}(j)$ whose corresponding map $\hat\gamma(\theta\sr(j),\cdot)\in\cM$ is thus the best model relating $\win$ and $\wout$ according to \eqref{d:cost_functional}. 

The main difficulty in the design of the identifier, is that the signals $\win$ and $\wout$ are \emph{not available for feedback}. In turn, in the overall regulator \eqref{s:ctr} the identifier is fed with the input $(\eta,u)$ in place of $(\win,\wout)$. In this way, $\eta$ plays the role of a ``proxy variable'' for $\win$, and $u$ for $\wout$ (notice that $\eta=\tau(w)=\win$ and $u=u\sr(w)=\wout$ in the ideal steady state in which $\y=0$). In turn, this is equivalent to provide the identifier with the ``corrupted input'' 
\begin{align*}
&\win+\din & &\wout+\dout  
\end{align*}
in which $\din = \eta-\win$ and $\dout=u-\wout$, and the resulting interconnection between the identifier and the core process \eqref{s:core_process}-\eqref{s:core_process_out} reads as follows.
\begin{equation}\label{s:core_identifier}
\begin{aligned}
&\left\{\begin{array}{lcl}
\dot\tim &=&1\\
\dot w &=& s(w)\\
\dot \xi &=& 0,\quad \dot\theta=0
\end{array}\right. \\ &\qquad(\tim,w,\xi,\theta,\din,\dout)\in \rC_\tim\x \rW\x\Xi\x\Theta\x\R^\deta\x\R^\dy\\ 
&\left\{ \begin{array}{lcl}
\tim^+ &=&0\\
w^+ &=& w\\
\xi^+ &=& \vhi(\xi, \win(w)+\din, \wout(w)+\dout)\\
\theta^+&=& \Thmap(\xi)
\end{array}\right.\\&\qquad(\tim,w,\xi,\theta,\din,\dout)\in \rD_\tim\x \rW\x\Xi\x\Theta\x\R^\deta\x\R^\dy,
\end{aligned}
\end{equation}
The choice of the remaining degrees of freedom $(\Xi,\vhi,\Thmap)$ is then made to satisfy a set of robust (with respect to $\din$ and $\dout$) stability and optimality conditions relative to the cost functional \eqref{d:cost_functional}. This conditions are formally expressed within the  forthcoming requirement, in which we make reference to the interconnection \eqref{s:core_identifier} where, for the sake of generality, the disturbance   $(\din,\dout)\in \R^\deta\x\R^\dy$ is treated as a generic exogenous input.

\begin{definition}	\label{def:identifierRequirement}
	The tuple $(\cM,\Xi,\vhi,\Theta,\Thmap)$ is said to satisfy the {\em identifier requirement} relative to $\cJ$, if there exist $\beta_{\xi}\in\cK\cL$, locally Lipschitz $\rho_\xi,\rho_\theta\in\cK$, a compact set $\Xi\sr\subset\Xi$ and, for each solution pair $((\tim,w,\xi,\theta),(\din,\dout))$ to \eqref{s:core_identifier}, a pair $(\xi\sr,\theta\sr):\dom(\xi,\theta)\to\Xi\x\Theta$ and a $j\sr\in\N$, such that $((\tim,w,\xi\sr,\theta\sr),(0,0))$ is a solution pair to \eqref{s:core_identifier} satisfying $\xi\sr(j)\in\Xi\sr$ for all $j\ge j\sr$, and  the following properties hold:
	\begin{enumerate}
		\item \label{req_opt} \textbf{Optimality:} for each $j\ge j\sr$
		\begin{equation*}
		\theta\sr(j)\in \optmap_{(\tim,w)}(j).
		\end{equation*}
		\item \label{req_stab}\textbf{Stability:} for each  $j\in\jumps(\tim,w)$  
		\begin{align*}
		|&\xi(j)-\xi\sr(j)| \\&\le\max\big\{ \beta_{\xi}(|\xi(0)-\xi\sr(0)|,j), \rho_\xi\left( |(\din,\dout)|_j\right)\big\}
		\end{align*}
		\item \label{req_reg} \textbf{Regularity:} 
		The function $\Thmap$ satisfies
		\begin{equation*}
		|\Thmap(\xi)-\Thmap(\xi\sr)| \le \rho_\theta(|\xi-\xi\sr|)
		\end{equation*}
		for all $(\xi,\xi\sr)\in\Xi\x\Xi\sr$,  the map 
		$(\theta,\eta)\mapsto \hat\gamma(\theta,\eta)$ is $\cC^1$ in the argument $\eta$, and $\partial \hat\gamma/\partial\eta$ is locally Lipschitz. \qedenv
	\end{enumerate}
\end{definition}
Examples of identifiers that fulfill these conditions 
are given in Section \ref{sec:DesignOfIdentifiers}.
The identifier requirement asks for the existence of a steady state $\xi\sr$ for the identifier such that the corresponding output $\theta\sr$ is optimal relative to \eqref{d:cost_functional} (optimality item). The optimal steady state $\xi\sr$ is required to be a solution to \eqref{s:core_identifier} whenever $(\din,\dout)=0$, i.e. when the identifier is fed by the ideal inputs $(\win,\wout)$, and it is required to be robustly stable when $(\din,\dout)$ is present (stability item).

Given a tuple $(\cM,\Xi,\vhi,\Theta,\Thmap)$ fulfilling the identifier requirement relative to a given cost functional $\cJ$, and with $\vep$  given by \eqref{d:vep}, with each solution pair $((\tim,w,\xi,\theta),(\din,\dout))$ of \eqref{s:core_identifier} we associate the \emph{optimal prediction error} 
\begin{equation}\label{d:varepsilon_star}
\vep\sr(w) := \vep(\theta\sr,w),
\end{equation}
 which represents the prediction error attained by the optimal model in the model set of the identifier computed along the ideal input-output data pair $(\win,\wout)=(\tau(w),u\sr(w))$.
\begin{remark}\label{rmk.id}
We stress that the signals $\win$ and $\wout$ \emph{are not assumed to be measured}, nor  they  are  used in the regulator~\eqref{s:ctr}. They have the sole role of posing a well-defined optimization problem, for which they  serve as ``nominal'' data set,  leading to a set of well-defined sufficient conditions for the design of the identifier (the identifier requirement). 
\end{remark}


\subsection{The Extended Observer}\label{sec:observer}
In this section we detail the choice of the degrees of freedom $(\Lambda(\ell),H,\ell,H_{r+1})$ characterizing the extended observer subsystem $(\hat x,\hat\sigma)$ of \eqref{s:ctr}, thus concluding the design of the regulator. The scalar $\ell$ is a positive control parameter that has to be taken large enough to ensure closed-loop stability, and it will be fixed in the forthcoming Theorem \ref{thm:main}. The matrix $\Lambda(\ell)$ is   chosen as $\Lambda(\ell) := \diag(\ell I_\dy,\,\ell^2 I_\dy,\,\dots,\,\ell^r I_\dy)$. 
For each $i=1,\dots,r+1$ and $j=1,\dots,\dy$, let $h_i^j\in\R$ be such that, for each $j=1,\dots,\dy$, the roots of the polynomials $\lambda^{r+1}+h^j_1 \lambda^r +\cdots + h^j_r\lambda + h^j_{r+1}$ 
are all real and negative. Then, the matrices $H$ and $H_{r+1}$ are defined as follows
\begin{equation*}
\begin{aligned}
H &:= \diag(H_1,\dots, H_r), & H_i&:= \diag(h^1_i,\dots,h^\dy_i)\\
H_{r+1} &:= \diag(h^1_{r+1},\dots,h^\dy_{r+1}).
\end{aligned}
\end{equation*}
Finally, with $\Xi\sr$ given by the identifier requirement, we define $\Theta\sr=\Thmap(\Xi\sr)$  and we let $\cH\sr\subset\R^\deta$ and $\cU\sr\subset\R^\dy$ be any compact sets satisfying $\tau(\rW)\subseteq\cH\sr$ and $u\sr(\rW)\subseteq \cU\sr$.
%
Then, 
we let $\psi$ be any  continuous function satisfying\footnote{We observe that a function $\psi$ with such properties can be simply obtained by saturating the map $\partial \hat\gamma(\theta,\eta)/\partial \eta\left(F\eta+Gu\right)$ with a saturation level larger or equal to $\bar\psi$.}
\begin{equation*}
\psi(\theta,\eta,u)  =   \pd{\hat\gamma(\theta,\eta)}{\eta}\left(F\eta+Gu\right)
\end{equation*}
for all $(\theta,\eta,u)\in\Theta\sr\x\cH\sr\x\cU\sr$ and, for some $\bar\psi>0$,
\begin{equation}\label{e:psi_bound} 
 |\psi(\theta,\eta,u)| \le \bar\psi
\end{equation}  
for all $(\theta,\eta,u)\in\Theta\x\R^\deta\x\R^\dy$.



%
%

\section{Main Result}\label{sec:MainResult}
The closed-loop system, obtained by interconnecting the plant \eqref{s:plant}, \eqref{s:exo} with the regulator \eqref{s:ctr}-\eqref{s:ctr_u}, results in the following hybrid system
\begin{equation}\label{s:cl}
 \begin{aligned}
&\left\{\begin{array}{lcl}
\dot\tim &=& 1\\
\dot{w}&=& s(w) \\
\dot{z}&=& f(w,z,x) \\ 
\dot{x}&=& A x + B \big(q(w,z,x) + b(w,z,x) u\big)\\
\dot\eta &=& F\eta + G u \\
\dot{\hat{x}} &=&  A \hat{x} + B(\hat{\sigma} + \bb  u) + \Lambda(\ell) HC (x-\hat{x})\\
\dot{\hat{\sigma}} &=& -\bb  \psi(\theta,\eta,u)+\ell^{r+1} H_{r+1}C (x-\hat{x})\\
\dot \xi &=& 0 ,\quad 
\dot\theta  =  0
\end{array}\right.\\
&\qquad (\tim,w,z,x,\eta,\hat x,\hat \sigma,\xi,\theta) \in \rC \\[.2em]
&\left\{\begin{array}{lcl}
\tim^+ &=& 0\\
w^+ &=& w,\quad 
z^+  =  z,\quad\, 
x^+  =  x,\quad \\
\eta^+ &=& \eta  ,\quad \,
\hat{x}^+  =   \hat x,\quad 
\hat{\sigma}^+  =  \hat\sigma \\
\xi^+ &=& \vhi(\xi,\eta,u) \\
\theta^+ &= & \Thmap (\xi) 
\end{array}\right.\\  
&\qquad  (\tim,w,z,x,\eta,\hat x,\hat \sigma,\xi,\theta) \in \rD
\end{aligned}  
\end{equation}
with $u$ given by \eqref{s:ctr_u} and with flow and jump sets given by $\rC:=\rC_\tim\x\rW\x\R^\dz\x\R^\dx\x\R^\deta\x\R^\dx\x\R^\dy\x\Xi\x\Theta$ and $\rD:=\rD_\tim\x\rW\x\R^\dz\x\R^\dx\x\R^\deta\x\R^\dx\x\R^\dy\x\Xi\x\Theta$.

In the remainder of the paper we let
\begin{equation}\label{d:cO}
\begin{aligned}
\cO  :=  \big\{& (\tim,w,z,x,\eta,\hat x,\hat\sigma,\xi,\theta)\in \rC\cup\rD   \st (w,z,x,\eta)\in \cE,  \\&\quad 
\hat x=x,\, \hat\sigma = -\bb u\sr(w)
\big\},
\end{aligned}
\end{equation}
with $\cE$ the set defined in \eqref{d:cE}. Furthermore, for every solution $\xb:=(\tim,w,z,x,\eta,\hat x,\hat\sigma,\xi,\theta)$ of \eqref{s:cl}, and with $(\xi\sr,\theta\sr)$ the trajectory produced by the identifier requirement relative to the solution pair $((\tim,w,\xi,\theta),(0,0))$ to \eqref{s:core_identifier}, we let for convenience
\begin{equation}\label{d:dist_cOsr}
|\xb|_{\cO\sr} := \max\left\{ |\xb|_\cO,\, |\xi-\xi\sr|     \right\} .
\end{equation}
Then, the following theorem is the main result of the paper.

\begin{theorem}\label{thm:main}
Suppose that Assumptions A\ref{ass:regularity}, A\ref{ass:minPhase} and A\ref{ass:b} hold, and let $\rZ_0\subset\R^\dz$ and $\rX_0\subset\R^\dx$ be arbitrary compact subsets. Consider the regulator \eqref{s:ctr}-\eqref{s:ctr_u}
{constructed in} Section \ref{sec:regulator}. Then the following holds:
\begin{enumerate}
	\item For each compact set $\rS_0\subset\R^\dx\x\R^\dy$ of initial conditions for $(\hat x,\hat\sigma)$, there exists $\ell\sr_{\rm s}>0$ such that, if $\ell\ge\ell_{\rm s}\sr$, then for every solution of the closed-loop system \eqref{s:cl} originating in $\Xb_0:=  (\rC_\tim\cup\rD_\tim)\x\rW\x\rZ_0\x\rX_0\x\R^\deta\x\rS_0\x\Xi\x\Theta$, $(w,z,x,\eta,\hat x,\hat\sigma)$ is  bounded    and satisfies $(w(t),z(t),x(t))\in\Omega_0$ for all $t\ge 0$, with $\Omega_0$ given in \eqref{e:Omega_0}. 
	\item  In addition, for each compact set $\rS_0$ and each  $\epsilon,T>0$, there exists $\ell_{\rm p}\sr(\epsilon,T)\ge \ell_{\rm s}\sr$ such that, if $\ell\ge\ell_{\rm p}\sr(\epsilon,T)$, then the trajectories $\xb$ of the closed-loop system \eqref{s:cl} originating in $\Xb_0$ also satisfy 
	\begin{equation}\label{e:thm_claim_21}
	\begin{aligned}
	|\hat x-x| &\le \epsilon, & |q(w,z,y)+b(w,z,y)u-\kappa(x)|&\le \epsilon
	\end{aligned} 
	\end{equation}
 for all $t\ge T$, and
	\begin{equation}\label{e:thm_claim_22}
	\limsup_{t+j\to\infty}|\xb(t,j)|_{\cO}\le \epsilon.
	\end{equation}
	
	\item If in addition $(\cM,\Xi,\vhi,\Theta,\Thmap)$ satisfies the identifier requirement relative to a cost functional $\cJ$, then for every solution $\xb$ of the closed-loop system \eqref{s:cl}, also $(\xi,\theta)$ is bounded, and there exists $\alpha_\xb >0$  and, for each compact set $\rS_0$ and each $\unT>0$, an $\ell\sr_{\rm \vep}(\unT)>\ell\sr_{\rm s}$, such that if $\ell>\ell\sr_{\rm \vep}(\unT)$ then
	\begin{equation}\label{e:thm_claim}  
	\limsup_{t+j\to\infty }|\xb(t,j)|_{\cO\sr}\le  \alpha_\xb \limsup_{t+j\to\infty} |\vep\sr(t,j)| ,
	\end{equation}
in which $\vep\sr(t,j):=\vep(\theta\sr(t,j),w(t,j))$.\qedenv
\end{enumerate} 
\end{theorem}

Theorem \ref{thm:main} is proved in Section \ref{sec:pf_thm}.
The first claim of the theorem is a boundedness result stating that if the observer gain $\ell$ is taken large enough, then all the trajectories originating in the chosen set of initial conditions are bounded and they have a common asymptotic bound. The second claim is a \emph{practical regulation} result extending that of \cite{Freidovich2008} and stating that, no matter how wrong the internal model and/or the identifier are, arbitrarily small error is eventually achieved by tuning the gain accordingly. The third claim is instead an \emph{approximate regulation} result relating the identifier prediction capabilities evaluated along the ideal data $(\win,\wout)=(\tau(w),u\sr(w))$ to the regulation performances in terms of asymptotic bound on the regulated variables. In particular, \eqref{e:thm_claim} implies that
	\begin{align*} 
	&\limsup_{t\to\infty }|\y(t)|\le  \alpha_\xb \limsup_{t+j\to\infty} |\vep\sr(t,j)|\\
	&	\limsup_{j\to\infty }|\theta(j)-\theta\sr(j)|\le  \alpha_\theta\Big( \limsup_{t+j\to\infty} |\vep\sr(t,j)|\Big) ,
	\end{align*}	
with $\alpha_\theta=\rho_\theta\circ\alpha_\xb$, which
 explicitly express  the asymptotic bound of the regulated variable $\y$ and the parameter estimation error $\theta-\theta\sr$ in terms of the optimal prediction error. Hence, as a consequence of the third claim, we also conclude that, whenever $\vep\sr=0$, i.e. when the actual internal model belongs to the identifier model set, then \emph{asymptotic regulation} and \emph{asymptotic parameter estimation} are achieved, thus extending the existence result of \cite{WanIsiLiuSu} to the adaptive case.
 
 \begin{remark}
 	In summary, the degrees of freedom  of the regulator \eqref{s:ctr} that have to be designed are: (i) the clock's upper and lower bounds $\unT$ and $\upT$ (Section \ref{sec:clock}); (ii) the stabilization and   saturation functions  $\kappa$ and $\sat$,  designed to robustly stabilize system \eqref{s:x_kappa} (Section \ref{sec:stabilizer}); (iii) the internal model pair $(F,G)$,  taken, according to \cite{Marconi2007}, so as $F$ is Hurwitz and $(F,G)$ controllable   (Section \ref{sec:internalmodel}); (iv) the identifier data $(\cM,\Xi,\vhi,\Theta,\Thmap)$,  chosen to satisfy the identifier requirement  (Section \ref{sec:Identifier}); (v) the extended observer data $(\Lambda(\ell),H,\ell,H_{r+1},\psi)$, fixed by following \cite{Freidovich2008} with an additional ``consistency term'' $\psi$ which is designed as a saturated version of the derivative of the identified model $\hat\gamma$ (Section \ref{sec:observer}). The control gain $\ell$, which needs to be chosen sufficiently large according to Theorem \ref{thm:main},  and depends on all the other quantities.
 \end{remark}

 \begin{remark}
We underline that the choice of the map $\kappa$ and $\sat$ detailed in Section \ref{sec:stabilizer} are independent from the observer, the internal model and the identifier. Besides, the result is global in $(\eta,\xi)$, which differs from \cite{WanIsiLiuSu} where the result is semi-global with respect to $\eta$, with the saturation in the controller that must be adapted to the initialization compact set of the internal model. On the other hand, the result is semi-global with respect to the observer, since the gain $\ell$ must be adapted to the observer initialization set $\rS_0$. 	\qedenv
 \end{remark}

 \begin{remark}\label{rmk:Holder}
  Assumptions A\ref{ass:regularity}-A\ref{ass:minPhase}, the consequent claim of Lemma \ref{lem:SICON}, and the identifier requirement in Definition \ref{def:identifierRequirement} all ask or state some Lipschitz  conditions on maps that play primary roles in the stability analysis. Nevertheless, we observe that all these regularity conditions may be relaxed to less restrictive H\"{o}lder continuity requirements by substituting the high-gain-based extended observer presented in Section \ref{sec:observer} with an \emph{homogeneous observer} of appropriate degree. The reader is referred to \cite{Andrieu2008} for further details.\qedenv 
 \end{remark}

%
%
\section{On the Design of   Identifiers}\label{sec:DesignOfIdentifiers}
\subsection{Least-Squares Identifiers for Linear Parametrizations}
\label{sec:leastsquares}

In this section we present a construction of the identifier when the model set $\cM$ consists of functions $\hat\gamma$ that are linear in the parameters $\theta$ and the cost functional \eqref{d:cost_functional} is a (weighted) \emph{least-squares} norm of the past prediction errors. For ease of notation  we focus here on the single-variable case (i.e. with $\dy=1$), as a multivariable identifier  can be obtained by concatenating of $\dy$ single-variable identifiers.
We consider a model set $\cM$ containing functions of the form
\begin{equation*}
 \hat\gamma(\theta,\cdot)  := \sum_{i=1}^{\dth} \theta_i \sigma_i(\cdot) = \theta^\top \sigma(\cdot),
\end{equation*}
with $\dth \in\N$ arbitrary, $\theta=\col(\theta_1,\dots,\theta_\dth)\in\R^\dth$, and $\sigma=\col(\sigma_1,\dots,\sigma_\dth)$, with $\sigma_i:\R^\deta\to\R$ differentiable functions with  locally Lipschitz derivative. The ``least-squares'' cost-functional is obtained by letting in \eqref{d:cost_functional} $\costf(s,i,j):=\mu^{j-i-1}|s|^2$, with $\mu\in(0,1)$ a design parameter playing the role of a \emph{forgetting factor}, and $\regf(\theta):=\theta^\top \Omega\theta$, in which $\Omega\in\SPD_\dth$. Thus, $\cJ$ reads as
\begin{equation}\label{d:ls_J}
\cJ_{(\tim,w)} (j)(\theta) := \sum_{i=0}^{j-1}\mu^{j-i-1}|\vep(\theta,w(t^i,i))|^2 + \theta^\top \Omega\theta.
\end{equation}
We design an identifier satisfying the identifier requirement relative to \eqref{d:ls_J} as follows. First, we let $\Theta:=\R^\dth$ and $\Xi:=\SPD_\dth \x \R^\dth$.  For a $\xi\in\Xi$ we consider the partition $\xi=(\xi_1,\xi_2)$ with $\xi_1\in\SPD_\dth$ and $\xi_2\in\R^\dth$, and we equip $\Xi$ with the norm $|\xi|:=|\xi_1|+|\xi_2|$. We consider the following \emph{persistence of excitation condition}, in which  we let $\msv(\cdot)$ denote the minimum non-zero singular value.
\begin{ass}\label{ass:PE}
	There exists   $\epsilon>0$ and, for each solution $(\tim,w)$ of the core process \eqref{s:core_process}, a $j\sr\in\N$, such that 
	\begin{equation}
	\msv\left( \Omega + \sum_{i=0}^{j-1}\mu^{j-i-1}\sigma\big(\tau(w(t^i))\big)\sigma\big(\tau(w(t^i))\big)^\top \right) \ge \epsilon.
	\end{equation}
	for all $j\ge j\sr$.\qedenv
\end{ass}
With $\cH\sr\subset\R^{\deta}$ and $\cU\sr\subset\R^\du$  compact subsets such that $\tau(\rW)\subseteq\cH\sr$ and $u\sr(\rW)\subseteq\cU\sr$, let  
\begin{align*} 
\rho_1 &:=  (1-\mu)\inv \sup_{\eta\in\cH\sr} |\sigma(\eta)\sigma(\eta)^\top|,\\
\rho_2 &:=   (1-\mu)\inv \sup_{(\eta,u)\in\cH\sr\x\cU\sr} |\sigma(\eta)u |,\\
\Xi\sr &:=  \Big\{ \xi\in\Xi\st \msv(\Omega+\xi_1)\ge \epsilon,\ |\xi_1|\le \rho_1,\ |\xi_2|\le \rho_2  \Big\}. 
\end{align*}  
Then,   with $\cdot\pinv$ denoting the Moore-Penrose pseudoinverse,   we let $\Sigma:\R^\deta\to\SPD_\dth$, $\lambda:\R^\deta\x\R^\du\to\R^\dth$ and $\Thmap:\Xi\to \R^\dth$ be any uniformly continuous functions satisfying
\begin{equation*}
\begin{array}{lcl}
\Sigma(\eta) &=& \sigma(\eta)\sigma(\eta)^\top\\
\lambda(\eta,u) &=& \sigma(\eta) u\\
\Thmap(\xi) &=& (\xi_1 + \Omega )\pinv \xi_2 
\end{array}
\end{equation*}
respectively on the compact sets $\cH\sr$, $\cH\sr\x\cU\sr$ and $\Xi\sr$, and
\begin{equation*}
|\Sigma(\eta)|\le \rho_\Sigma,\quad |\lambda(\eta,u)|\le \rho_\lambda,\quad |\Thmap(\xi)|\le c_{\Thmap}
\end{equation*}
everywhere else, for some $\rho_\Sigma,\ \rho_\lambda,\ \rho_\Thmap >0$.  Then the identifier is described by the following equations
\begin{equation}\label{s:ls}
\begin{array}{lcl}
\xi_1^+ &=& \mu\xi_1 + \Sigma(\win)\\
\xi_2^+ &=& \mu\xi_2 + \lambda(\win,\wout)\\
\theta^+ &=& \Thmap(\xi),
\end{array}
\end{equation}
and the following result holds.
\begin{proposition}\label{prop:ls}
	Assume A\ref{ass:PE}. Then, the identifier \eqref{s:ls} satisfies the identifier requirement relative to \eqref{d:ls_J}.\qedenv
\end{proposition}

The proof of Proposition \ref{prop:ls} can be deduced by the same arguments of \cite{Bin2018} and it is thus omitted.
 It is worth observing that, whenever the regularization matrix $\Omega$ is positive definite, A\ref{ass:PE} \emph{always holds} with $j=0$ and  $\epsilon$  the smallest eigenvalue\footnote{ 
 		To see this, pick any $M\in\SPD_\dth$ and  let  $\nu$ be an eigenvalue of $M+\Omega$  and $e$ an associated eigenvector. Then, $\epsilon|e|^2\le e^\top(M+\Omega)e = \nu |e|^2$, and thus $\nu\ge\epsilon$. This, in turn shows that $M+\Omega>0$   and $\msv(M+\Omega)\ge \epsilon$. Thus, the claim follows by noticing that the sum appearing in A\ref{ass:PE} is in $\SPD_\dth$.} of $\Omega$.
  The importance of  regularization is well understood in system identification (see e.g. \cite{Sjoberg1993}), although it is also well-known that it introduces a bias on the parameter estimation, in the sense that in   case a ``true map'' $\gamma$ relating $\win$ and $\wout$ exists and  belongs to   $\cM$,   the ``true parameter'' $\theta\sr$ is a minimum of \eqref{d:ls_J} only if $\theta\sr\in\ker\Omega$, so that having $\Omega$ nonsingular makes the identifier \eqref{s:ls}  converge ``only'' to a neighborhood of $\theta\sr$ whose size is related to the eigenvalues of $\Omega$ (and thus can be made arbitrarily small). Therefore, the regularization matrix $\Omega$ is a degree of freedom that must be chosen to weight well-conditioning of the problem and asymptotic estimation performances.
 If  $\Omega$ is chosen singular (possibly the zero matrix), the identifier requirement is still satisfied along the trajectories of $w$ that are persistently exciting according to A\ref{ass:PE}. In this respect, we observe that A\ref{ass:PE} is a property of the ideal input signal $\win=\tau(w)$ and of chosen clock strategy, as the sampling time of the core process \eqref{s:core_process} depends on it.

\subsection{``Mini-Batch'' Algorithms for Nonlinear Parametrizations}\label{sec:minibatch}
In this section we present a construction of the identifier fulfilling the identifier requirement when the model set $\cM$ assumes the generic form \eqref{d:model_set},  with $\Theta=\R^\dth$ for some $\dth\in\N$, and with $\eta\mapsto\hat\gamma(\theta,\eta)$ which is $\cC^1$ for all $\theta$ and such that $\partial\hat\gamma/\partial\eta$ is locally Lipschitz. We start by assuming to have available a   \emph{batch identification algorithm} working on a data set of finite size $N$, and we define an identifier fitting in our framework that repeatedly executes the algorithm on a ``moving window'' of   size $N$.

More precisely, with $\Sn_n$ the space of functions $\{1,\dots, N\}\to\R^n$, for any two signals $s_{\rm in}\in\Sn_\deta$ and $s_{\rm out}\in \Sn_\dy$  we define the \emph{window cost}   
\begin{equation*}
\cI^N_{(s_{\rm in},s_{\rm out})} (\theta) := \sum_{i=1}^N \varpi\big( s_{\rm out}(i) - \hat\gamma(\theta,s_{\rm in}(i)), i\big)  + \rho(\theta),
\end{equation*}
for some integral cost $\varpi:\R^\dy\x\{1,\dots,N\}\to\Rplus$ and regularization term $\rho:\R^\dth\to\Rplus$. 

Then we assume the following.
\begin{ass}\label{ass:minibatch}
There exists a Lipschitz map $\cG:\Sn_\deta\x\Sn_\dy\to\R^\dth$ such that, for every solution $(\tim,w)$ to the core process \eqref{s:core_process} and for every $j\ge N$,   with 
\begin{align*}
s_{\rm in}^j(i) &:= \tau\big(w(t^{j+i-N-1})\big), & s_{\rm out}^j(i) &:= u\sr\big(w(t^{j+i-N-1})\big),
\end{align*}
for  all $i=1,\dots, N$, it holds that
\begin{equation*}
\cG(s_{\rm in}^j,s_{\rm out}^j) \in \argmin{\theta\in\R^\dth} \cI^N_{(s_{\rm in}^j,s_{\rm out}^j)}(\theta) .
\end{equation*}\qedenv
\end{ass}
The map $\cG$ represents any optimization algorithm that extracts the optimal model of $\cM$ from the finite data set represented by the ``windowed samples'' $(s_{\rm in},s_{\rm out})$ of $(\win,\wout)=(\tau(w),u\sr(w))$. With $\lambda_n:\R^{Nn}\to\Sn_n$ the linear operator mapping the vector $v=(v_1,\dots,v_N)$, $v_i\in\R^n$, to the signal $s\in\Sn_n$ satisfying $s(i):=v_i$, we construct an identifier starting from $\cG$ by letting $\Theta:=\R^\dth$, $\Xi:=\R^{N\deta}\x\R^{N\dy}$, and $(\vhi,\Thmap)$ such that the state $\xi:=(\xi_1,\xi_2)$, with $\xi_1\in\R^{N\deta}$ and $\xi_2\in\R^{N\dy}$, and output $\theta$ of the identifier satisfy
\begin{equation}\label{s:mb_xi}
\begin{array}{lcl}
\xi_1^+ &=& H_1 \xi_1 + B_1 \win\\
\xi_2^+ &=& H_2 \xi_2 + B_2 \wout \\
\theta^+&=& \Thmap(\xi),
\end{array}
\end{equation}
in which, for $i=1,2$, $(H_i,B_i)$ have the ``shift'' form
\begin{equation*}
H_i :=\begin{pmatrix}
0_{N m_i\x m_i}\vline & \begin{matrix}
I_{(N-1) m_i}\\ 0_{ m_i\x(N-1)\dy)}
\end{matrix} 
\end{pmatrix},  B_i :=\begin{pmatrix}
0_{(N-1) m_i\x  m_i }\\ I_{m_i} 
\end{pmatrix} 
\end{equation*}
where we let $m_1=\deta$ and $m_2=\dy$, and
\begin{equation}\label{s:mb_thmap}
\Thmap(\xi) := \cG(\lambda_{\deta}(\xi_1),\lambda_\dy(\xi_2)).
\end{equation}
The identifier \eqref{s:mb_xi} consists of a pair of ``shift registers'' propagating and accumulating the new values of $(\win,\wout)=(\tau(w),u\sr(w))$ and forming in this way a moving window. The output map \eqref{s:mb_thmap} assigns to the parameter $\theta$ the value given by the algorithm $\cG$ corresponding to the current data set stored in the state $\xi$. This construction has the following property, proved in Appendix \ref{apd:proof_minibatch}.
\begin{proposition}\label{prop:minibatch}
Assume A\ref{ass:minibatch}, then the identifier \eqref{s:mb_xi}-\eqref{s:mb_thmap} satisfies the identifier requirement relative to the cost functional
\begin{equation*}
\cJ_{(\tim,w)}(j)(\theta) :=\!\! \sum_{i=\max\{0,j-N\}}^{j-1}\!\!\! \varpi\big( \vep(\theta,w(t^i)) ,i-j+N+1 \big) + \rho(\theta) 
\end{equation*}
and with $j\sr=N$ for each solution $(\tim,w)$ of  \eqref{s:core_process}.\qedenv
\end{proposition}

\section{Example}\label{sec:Example}
We consider the problem of synchronizing the output of a Van der Pol oscillator with \textit{unknown} parameter, with a triangular wave with \emph{unknown} frequency. The plant, which consists in a forced  Van der Pol oscillator, is described by the following equations
\begin{equation}\label{ex:s:p}
\begin{array}{lcl}
\dot p_1 &=& p_2 \\ \dot p_2 &=& -p_1 + a(1-p_1^2)p_2 +   u ,
\end{array}
\end{equation}
	with $a$ an unknown parameter known to range in $[\underline{a},\overline{a}]$ for some constants $\bar a>\und a>0$.
According to \cite{LeeSmi}, a triangular wave can be  generated by an exosystem of the form
\begin{equation}\label{ex:w}
\begin{aligned} 
\dot w_1 &=  w_2, &
\dot w_2 &=  -\varrho w_1 
\end{aligned}
\end{equation}
with output
\begin{equation*}
p_1\sr(w) := 2\sqrt{w_1^2+w_2^2}\arcsin\left(\dfrac{w_1}{\sqrt{w_1^2+w_2^2}}\right),
\end{equation*}
and in which $\varrho$ is the unknown frequency, assumed to lie in the interval $[\und\varrho,\bar\varrho]$ with $\bar\varrho>\und\varrho>0$ known constants.
The control goal thus consists in driving the output $p_1$ of \eqref{ex:s:p} to the reference trajectory $p_1\sr(w)$. With $s(w):=(w_2,-\varrho w_1)$, we define the error system $x$ as 
\begin{equation*}
x := \begin{pmatrix}
x_1\\x_2
\end{pmatrix} = \begin{pmatrix}
p_1 - p_1\sr(w)\\
p_2- L_{s(w)} p_1\sr(w) 
\end{pmatrix},
\end{equation*}
which is of the form \eqref{s:plant}, without $z$, with 
\begin{align*}
\y &=  x_1 = p_1 - p_1\sr(w)\\
q(w,x) &:= -x_1 -p_1\sr(w)- L_{s(w)}^2 p_1\sr(w)\\
&\qquad + a\big(1-(x_1+p_1\sr(w))^2\big)\big(x_2+L_{s(w)}p_1\sr(w)\big),
\end{align*}
and with $b(w,x)  := 1$.
The ideal steady-state error-zeroing control action that the regulator should provide is given by
\begin{equation*}
\begin{aligned}
u\sr(w) &= -q(w,0)/b(w,0) \\&=  p_1\sr(w)+L_{s(w)}^2 p_1\sr(w)  - a(1-p_1\sr(w)^2)L_{s(w)}p_1\sr(w)  ,
\end{aligned}
\end{equation*}
and no analytic technique is known to compute the right function $\gamma$ of the internal model of \cite{Marconi2007,WanIsiLiuSu} for which the regulator is able to generate $u\sr(w)$. 

Regarding the exosystem \eqref{ex:w}, we observe that the quantity $V_\varrho (w_1,w_2) :=   \varrho w_1/2 +   w_2/2$ 
remains constant along each solution. Hence,  the set $W := \bigcup_{\varrho\in[\und\varrho,\bar\varrho]} V_{\varrho}\inv([0,c])$ 
is invariant for \eqref{ex:w}. Furthermore, assumptions A\ref{ass:regularity}, A\ref{ass:minPhase} and A\ref{ass:b} hold by construction, with $\bb=1$ and any $\mu\in(0,1)$, and hence, the problem fits into the framework of this paper, and the proposed regulator is used with:

\begin{enumerate}
	\item[(i)] $\kappa(x)=-Kx$, with $K\in\R^{2\x 2}$ such that $\sigma(A-BK)=\{-1,-2\}$, and $\sat$ implements the standard saturation function with level $\rM=100$;
	
	\item[(ii)] $\deta=2(\dw+1)=6$ and 
	\begin{align*}
	F&:=\begin{pmatrix}
	-1 & 1 & 0 & 0 & 0 & 0\\
	0& -1 & 1 & 0 & 0  & 0\\
	0 & 0 &-1 & 1 & 0 &  0\\
	0 & 0 & 0 &-1 & 1 &  0\\
	0& 0 & 0 & 0 &  -1 & 1 \\
	0& 0 & 0 & 0 & 0 &  -1 \\
	\end{pmatrix}, & G&:=\begin{pmatrix}
	0\\0\\0\\0\\0\\1
	\end{pmatrix};
	\end{align*}
	\item[(iii)] the identifier is chosen as a least-squares identifier of the kind presented in Section \ref{sec:leastsquares}, in which the regressor vector $\sigma$ is defined to perform a polynomial expansion of $\gamma$ with a polynomials of odd order. More precisely, with $N\in\N$, for $n\le N$ we let 
	\begin{equation*}
	\cI_n:=\big\{  (i_1,\dots,i_n)\in\{1,\dots,6\}^n \st i_1\le \dots\le i_n \big\}
	\end{equation*}
	be the set of non-repeating multi-indices of length $n$, and with $I\in\cI_n$, we let $\sigma_I(\eta) :=   \eta_{i_1}\cdot \ldots \cdot \eta_{i_n}$. 
	The regressor $\sigma$ is then defined as 
	\begin{equation*}
	\sigma = \col\big(\sigma_I\st I\in \cI_n,\ n\le N,\ n\text{ odd} \big).
	\end{equation*}
	In the forthcoming simulations we have taken $N=1,3,5$. To ensure that the persistence of excitation condition of A\ref{ass:PE} holds, we have taken a diagonal regularization matrix $\Omega=10^{-3} I$. The forgetting factor is instead chosen as $\mu=0.99$.

\item[(iv)] the extended observer is implemented with  $\ell=20$, $h_1=6$, $h_2=11$, $h_3=6$ and with $\phi$ that is obtained by saturating the function $(\theta,\eta,u)\mapsto (\partial\hat\gamma(\theta,\eta)/\partial\eta)(F\eta+Gu)$ with a saturation level of $100$. 
\item[(v)] finally, a periodic clock strategy is employed, obtained by letting $\unT=\upT=0.1$.
 
\end{enumerate}

\begin{figure}[t]
	\centering
	\includegraphics[width=\linewidth,trim=4em 2em 3.7em 2em,clip]{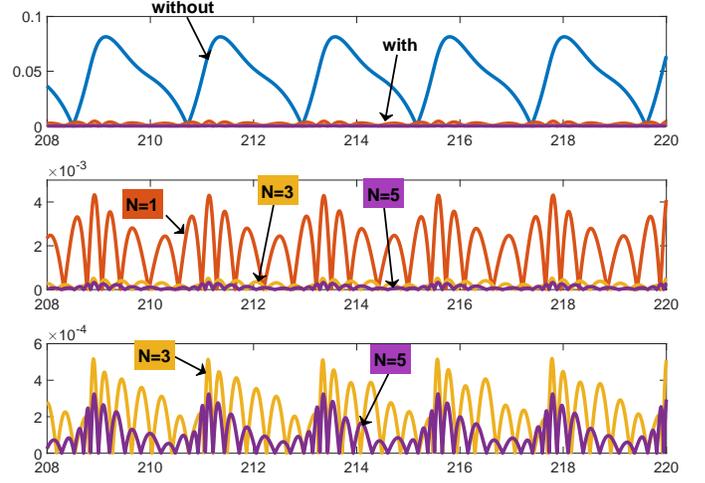}
	\caption{\textbf{Top:} steady-state time evolution of the tracking error $\y$ in the four cases obtained without adaptive internal model and with $N=1,3,5$. \textbf{Center} and \textbf{Bottom}: zoom-in to highlight the difference between the four cases. In abscissa:   time (in seconds).}
	\label{fig:err}\vspace{-1em}
\end{figure}
\begin{figure}[t]
	\centering
	\includegraphics[width=\linewidth,trim=5em .5em 3.7em 0em,clip]{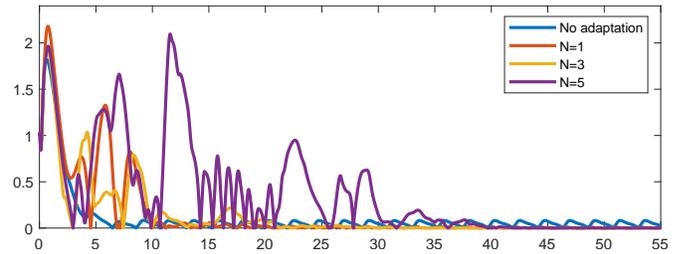}
	\caption{Transitory of the tracking error  $\y$ in the four cases obtained without adaptive internal model and with $N=1,3,5$. In abscissa:   time (in seconds).}
	\label{fig:trans}\vspace{-1em}
\end{figure}
\begin{figure}[t]
	\centering
	\includegraphics[width=\linewidth,trim=4em 1.0em 3.7em 0em,clip]{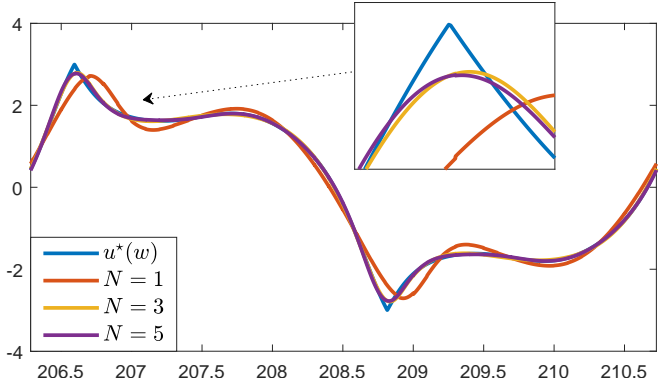}
	\caption{Time evolution of $u\sr(w(t))$ and of its approximation $\hat\gamma(\theta(t),\eta(t))$ for $N=1,3,5$. In abscissa:   time (in seconds).}
	\label{fig:gamma}\vspace{-1em}
\end{figure}

The following simulation shows the regulator applied with $a=\varrho=2$ in four cases: (1) without internal model\footnote{We stress that this is equivalent to implement an extended observer as that proposed in \cite{Freidovich2008} (see also Section~\ref{sec.framework.previous}).}, i.e. with $\phi=0$; (2) with the adaptive internal model obtained by setting $N=1$ (i.e., with $\sigma(\eta)=\eta$); (3) with $N=3$, i.e. with $\sigma(\eta)=\col(\eta_1,\dots,\eta_6,\eta_1^3,\eta_1^2\eta_2,\eta_1\eta_2\eta_3\dots,\eta_6^3)$ and (4) with $N=5$, i.e. with $\sigma(\eta)=\col(\eta_1,\dots,\eta_6,\eta_1^3,\dots,\eta_6^3,\eta_1^5,\dots,\eta_6^5)$. In particular, Figure \ref{fig:err} shows the steady-state evolution of the tracking error $\y(t)=p_1(t)-p_1\sr(w(t))$ in the four cases described above. The error obtained by introducing a linear adaptive internal model ($N=1$) is reduced by more than $15$ times compared to the case in which the adaptive internal model is not present (i.e. $\phi=0$). Adding to the model set the polynomials of order $3$ $(N=3)$,   reduces the maximum error of more than 120 times compared to the first case without internal model. Finally, with $N=5$, the maximum error is reduced by more than $200$ times.

Figure~\ref{fig:trans} shows the transitory of the tracking error trajectories corresponding to the four cases described above. It can be noticed that the settling time increases with the number $N$. This is due to the fact that an increase of  $N$ leads to an increase of the overall number of parameters, each one having its own transitory. From Figures~\ref{fig:err} and \ref{fig:trans} we can thus conclude that a larger model complexity is associated with a smaller steady-state error but with a longer transitory.

Finally, Figure \ref{fig:gamma} shows   the time evolution of the ideal steady-state control law $u\sr(w)$ and of its approximation given by $\hat\gamma(\theta(j),\eta(t))$ in the three cases in which $N=1,3,5$.

%
%
 \section{Proof of Theorem \ref{thm:main}}\label{sec:pf_thm}
We subdivide the proof in three parts, coherently with the three claims of the theorem.
For compactness, in the following we will  write  $\pb:=(w,z,x)$ in place of $(w,z,x)$, and we let
\begin{equation*} 
\fb(\pb,u)  := \col\big(s(w),f(\pb), Ax +B(q(\pb)+b(\pb)u) \big).  
\end{equation*} 
Moreover, we also use the symbol $\placeholder$ in place of the arguments of functions that are uniformly bounded (we refer in particular to \eqref{d:sat} and \eqref{e:psi_bound}).
\subsection{Stability analysis}\label{sec:proof:stability}
Since $\unT>0$ and $\upT<+\infty$, then all the complete trajectories of \eqref{s:cl} have an infinite number of jump  and flow times.
Since $\eta$ is a Hurwitz linear system driven by a bounded input $u$, its solutions are complete and bounded.  Regarding the subsystem $(w,z,x,\hat x,\hat \sigma)$, let
\begin{equation}\label{pf:d:sigma}
\begin{array}{lcl} 
\sigma  &:=& q(p)+ (b(p)-\bb)\u(x,\hat{\sigma}),\\
\u(x,\sigma) &:=& \bb\inv \sat(-\sigma +\kappa(x)  ).
\end{array}
\end{equation} 
In view of \eqref{s:ctr_u}, $u=\u(\hat x,\hat\sigma)$. Hence the dynamics of $x$ can be rewritten as 
\[
\dot{x} = A x + B\big(\sigma + \bb u + (b(p)-\bb)(\u(\hat{x},\hat{\sigma})-\u(x,\hat{\sigma})) \big)  .
\]
Following the standard high-gain paradigm, define
\begin{equation}\label{pf:e}
\begin{aligned}
\e_x &:= \ell\Lambda(\ell)\inv(\hat{x}-x)  , & e_\sigma &:= \ell^{-r}(\hat{\sigma}-\sigma)
\end{aligned}
\end{equation}
and change coordinates according to 
\begin{equation*}
(\hat x,\hat\sigma ) \mapsto \e := \col\big(  \e_x, \e_\sigma  \big).
\end{equation*}
In the new coordinates, \eqref{s:ctr_u} reads as
\begin{equation}\label{pf:phi_u}
u =  \bb\inv \sat(-\ell^r\e_\sigma -\sigma +\kappa(\Lambda(\ell)\ell\inv\e_x+x)  ),
\end{equation}
and \eqref{pf:d:sigma} gives rise to the implicit equation \begin{equation}\label{pf:eq_Tsigma}
T_\sigma(p, \e_\sigma ,\sigma) = 0,
\end{equation}  
where  $T_\sigma (p,\e_\sigma,\sigma):=\sigma- q(p)  - (b(p)-\bb)\bb\inv \sat(   -\ell^r \e_\sigma - \sigma + \kappa(x))$. 
We observe that
\begin{equation*}
\pd{T_\sigma}\sigma (p,e_\sigma,\sigma) =  I + (b(p)-\bb)\bb\inv  \sat'(\placeholder).
\end{equation*}
Thus, A\ref{ass:b} and \eqref{d:sat} give $|(b(p)-\bb)\bb\inv \sat'(\placeholder)|\le 1-\mu$, so that $\partial T_\sigma/\partial \sigma$ is uniformly nonsingular. This, in turn, suffices to show that there exists a unique $\cC^1$ function $\sol_\sigma(p,\e_\sigma)$  
satisfying $T_\sigma(p,\e_\sigma,\sol_\sigma(p,\e_\sigma))=0$, and such that  
\begin{equation}\label{pf:d_phi_sigma}
\sigma = \sol_\sigma(p,\e_\sigma).
\end{equation} 
We further notice that, $T_\sigma(p,\e_\sigma,\sigma)=0$ also implies 
\begin{align*}
\pd{T_\sigma}{p}(p,\e_\sigma,\sigma )\fb(p,u) \!+\! \pd{T_\sigma}{\e_\sigma}(p,\e_\sigma,\sigma) \dot\e_\sigma \!+\! \pd{T_\sigma}{\sigma}(p,\e_\sigma,\sigma)\dot\sigma \!=\! 0
\end{align*} 
which in turn yields
\begin{equation*}
\dot\sigma = \Delta_1  
+ \Delta_2 \ell^r\dot\e_\sigma  ,
\end{equation*}
where,
with
$m(p,\placeholder) := (b(p)-\bb)\bb\inv \sat'(\placeholder)$,
we let
\begin{equation*}
\begin{aligned}
\Delta_1 & = -(I+m(p,\placeholder))\inv\pd{T_\sigma}{p}(p,\e_\sigma,\sigma ) \fb(p,u), \\
\Delta_2 & = -(I+m(p,\placeholder))\inv m(p,\placeholder). 
\end{aligned}
\end{equation*}
Due to A\ref{ass:b} and \eqref{d:sat}, $|m(p,\placeholder)|\le 1-\mu$, so that $I+m(p,\placeholder)$ is always invertible. Moreover, in view of  \cite[Proposition 10.3.2]{CampbellPseudoInvBook}, $|(I+m(p,\placeholder))\inv|  \le \mu\inv$, so that  we obtain
\begin{align}\label{pf:bound_Delta32}
\Delta_2 & \le \mu\inv-1 
\end{align}
for all $(p,\e_\sigma,\sigma)\in\rW\x\R^\dz\x\R^\dx\x\R^\dy\x\R^\dy$.
The variable $\e$ jumps according to $\e^+=\e$ and flows according to
\begin{align}\label{pf:dot_e_x}
\dot\e_x &= \ell (A-HC)\e_x + B  (\ell \e_\sigma 
+ \Delta_{3,\ell} )
\\
\label{pf:dot_e_sigma}
\dot\e_\sigma &= -\ell H_{r+1} C\e_x - \Delta_2\dot\e_\sigma 
+ \ell^{-r} \Delta_4 
\end{align}
where $\Delta_{3,\ell}  :=  -
\ell^{1-r}	(b(p)-\bb)(\u(\hat{x},\hat{\sigma})-\u(x,\hat{\sigma}))$ and $\Delta_4  := -\bb \psi(\placeholder) - \Delta_1$. 
Since  
$
I+\Delta_2  = I-(I+m(p,\placeholder))\inv m(p,\placeholder) = (I+m(p,\placeholder))\inv
$
and $|m(p,\placeholder)|\le 1-\mu<1$, then $I+\Delta_2$ is uniformly invertible, and solving \eqref{pf:dot_e_sigma} for $\dot\e_\sigma$ yields
\begin{equation*}
\dot\e_\sigma = -\ell H_{r+1}C\e_x - m(p,\placeholder)\ell H_{r+1}C\e_x +  \ell^{-r} \Delta_5 .
\end{equation*}
with $\Delta_5:=(I+m(p,\placeholder))\Delta_4$.
Hence, by letting $\e:=\col(\e_x,\e_\sigma)$ and
\begin{equation*}
\Ae := \begin{pmatrix}
A-HC & B\\
-H_{r+1}C & 0_\dy
\end{pmatrix}, \ 
  \Be_x:=\begin{pmatrix}
B \\ 0_{\dy} 
\end{pmatrix},\ 
 \Be_\sigma:=\begin{pmatrix}
0_{\dx\x\dy}\\ I_\dy 
\end{pmatrix} 
\end{equation*}
we obtain
\begin{equation*}
\dot \e = \ell \Ae \e + \Be_x\Delta_{3,\ell}+ \Be_\sigma \big( - m(p,\placeholder)\ell H_{r+1}C\e_x+ \ell^{-r}\Delta_5  \big)   .
\end{equation*}
Let $\Omega_0$ be the compact set introduced in Section \ref{sec:stabilizer} and fulfilling \eqref{e:Omega_0}. In view of A\ref{ass:regularity} and \eqref{d:sat}, there exists $a_0>0$  such that $|\Delta_1|\le a_0$ holds for all $p\in\Omega_0$ and all $(e_\sigma,\hat{\sigma})\in(\R^\dy)^2$. In view of  \eqref{e:psi_bound} and \eqref{pf:bound_Delta32}, $|\Delta_5| \le a_1$ with 
$a_1 := |\bb|\bar\psi+a_0$.
Moreover,  since $\kappa$ and $\sat$ are $\cC^1$, and $u$ is bounded, then it is $\cC^1$ and Lipschitz. Thus, A\ref{ass:b} and \eqref{pf:e} imply the existence of $a_2>0$ such that $|\Delta_{3,\ell}|\le   a_2 |\e_x|$ for all $\ell\geq 1$ and   all $(p,e_\sigma,\hat{\sigma})$. 
The stability properties of $\e$ then follow by the Lemma below,  proved in Appendix \ref{apd:proof_lemma_e_mu}.
\begin{lemma}\label{lem:e_mu}
	Consider a system of the form
	\begin{equation}\label{pf:chi}
	\dot \chi = \ell \Ae\chi -\ell\alpha(t)\Be_\sigma  H_{r+1} C\chi_1 + \Be_x  \delta_1+ \delta_2
	\end{equation}
	With $\chi=\col(\chi_1,\chi_2)$, $\chi_1\in\R^{\dx}$, $\chi_2\in\R^{\dy}$,    $\delta_1:\Rplus\to\R^\dy$ and  $\delta_2:\Rplus\to\R^{\dx+\dy}$  locally integrable inputs such that, for some $\pi_1>0$, $|\delta_1|\le \pi_1|\chi|$,  and $\alpha:\Rplus\to\R$ a continuous function satisfying $|\alpha(t)|\le \bar\alpha<1$ for all $t\in\Rplus$. Then there exist $\ell\sr_0,\pi_2,\pi_3,\pi_4>0$ such that, if $\ell\ge\ell\sr_0$, then
	\begin{equation*}
	|\chi(t)|\le \max\Big\{\pi_2 \ee^{-\pi_3\ell t}|\chi(0)|, \, \pi_4\ell\inv |\delta_2|_t \Big\}
	\end{equation*}
 for all $t\in\Rplus$ for which the solution is defined.\qedenv
\end{lemma}

In particular, since $|m(p,\placeholder)|\le 1-\mu<1$,  Lemma \ref{lem:e_mu} yields the existence of $\ell\sr_0,a_3,a_4,a_5>0$ such that, if $\ell\ge\ell\sr_0$, then
\begin{equation}\label{pf:e_bound}
 	|\e(t)|\le \max\Big\{a_3\ee^{-a_4 \ell t}|\e(0)|, \, a_5\ell^{-(r+1)} |\Delta_5|_t \Big\}.
\end{equation} 
Moreover, the following Lemma holds.
\begin{lemma}\label{lem:ell_epsilon}
Suppose that, for some $T_0,a_1>0$, $|\Delta_5|_t\le a_1$ for all $t\in[0,T_0)$. Then, for each $T\in(0,T_0)$ and each $\epsilon>0$, there exists $\ell\sr_1(T,\epsilon)\ge\ell\sr_0$ such that, if $\ell\ge\ell\sr_1(T,\epsilon)$, then for each solution of \eqref{s:cl} originating in $\Xb_0$, it holds that
\begin{equation*}
\max\{|x(t)-\hat x(t)|,|\sigma(t)-\hat\sigma(t)|\}\le \epsilon 
\end{equation*}
for all $t\in[T,T_0)$.\qedenv
\end{lemma}

Lemma \ref{lem:ell_epsilon} is proved in Appendix \ref{apd:proof_lemma_ell_epsilon}.
Let 
\begin{equation*}
b_1  :=\sup_{(p,s)\in\Omega_0\x\R^\dy} \left|Ax+B\big(q(p)+b(p)\bb\inv\sat(s)\big)\right| .
\end{equation*} 
Then,  as long as $p\in\Omega_0$, we have $|\dot x|\le b_1$. Thus,  for all $t\in\Rplus$ such that $p(t)\in\Omega_0$, it holds that $ |x(t)|\le |x(0)| + b_1t$. 
In view of \eqref{e:Omega_0},   $|x(0)|\le \varrho_0$, so that $|x(t)|\le \varrho_1$ for all $t\le \bar t_1:=(\varrho_1-\varrho_0)/b_1$. Furthermore, \eqref{e:Omega_0} also yields  $|w(0),z(0)|_\cA\le \varrho_0$. As A\ref{ass:minPhase} implies that $(w,z)$ exhibits no finite escape time in $[0,\bar t_1]$, then by continuity there exists $\bar t \in(0,\bar t_1)$ such that $\max\{|x(t)|,\,|(w(t),z(t))|_\cA\}\le \varrho_1$, for all  $t\in[0,\bar t]$. 
This, with \eqref{e:Omega_0}, implies that $p(t)\in \closed\ball_{\varrho_1}^\cB\subset\Omega_0$ for all $t\in[0,\bar t]$ along the trajectories originating inside $\Xb_0$.
%
\begin{lemma}\label{lem:sigma}
 The unique solution \eqref{pf:d_phi_sigma} of \eqref{pf:eq_Tsigma} satisfies
 \begin{equation}\label{pf:phi_sigma_lemma}
 \phi_\sigma(p,\e_\sigma) =  \bb b(p)\inv q(p) + (I-\bb b(p)\inv) \big( \kappa(x)-\ell^r\e_\sigma \big) 
 \end{equation} 
 for all $(p,\e_\sigma)\in\Omega_0\x\R^{\dy}$ such that  $|\ell^r\e_\sigma|\le \mu \varrho_2$,
  with $\mu$ and $\varrho_2$ the constants given respectively in A\ref{ass:b} and \eqref{d:satlev}.\qedenv
\end{lemma}

Lemma \ref{lem:sigma} is proved in Appendix \ref{apd:proof_lemma_sigma}. Pick $\nu>0$ and let 
\[
\rX_{\nu} = \big\{ \hat{x}\in \R^\dx \st  |x-\hat{x}|\leq \nu,\   (w,z,x)\in \Omega_0  \big\}  .
\]
Then $\rX_{\nu}$ is compact and, by continuity of $\kappa$, there exists $\rho_\kappa\in\cK$ such that $
|\kappa(\hat{x})-\kappa(x)|\le \rho_\kappa(|\hat{x}-x|) 
$  for all $(\hat{x},x)\in (\rX_{\nu})^2$.

%
With $\varrho_2$ and $\bar\delta$ defined in Section \ref{sec:stabilizer}, let $\ell\sr_{\rm s}$ be taken equal to the $\ell\sr_1(T,\epsilon)$ produced by Lemma~\ref{lem:ell_epsilon} for $T \in(0,\bar t)$ and 
\begin{equation}\label{pf:epsilon}
\epsilon=\min\left\{\nu,\ \mu\varrho_2/2,\ \rho_\kappa\inv( \varrho_2/2) ,\ \rho_\kappa\inv(\mu \bar\delta/2),\ \bar\delta/2 \right\},
\end{equation}
and pick $\ell>\ell\sr_{\rm s}$. As $p(t)\in\Omega_0$ in $[T,\bar t)$, then Lemma~\ref{lem:ell_epsilon} and \eqref{pf:epsilon} imply $|\ell^r\e_\sigma|\le\mu\varrho_2$ in $[T,\bar t)$. Thus,  by Lemma \ref{lem:sigma}   $\sigma(t)=\phi_\sigma(p(t),\e(t))$ satisfies \eqref{pf:phi_sigma_lemma} in $[T,\bar t)$.  Moreover, for   $t\in[T,\bar t)$, $\hat x(t)\in\rX_{\nu}$  and the argument of   $\sat$  in \eqref{pf:phi_u}   satisfies
\begin{align*}
|&-\sigma - \ell^r\e_\sigma + \kappa(\Lambda(\ell)\ell\inv\e_x+x)|\\
&\le \max_{p\in\Omega_0}|\bb c(p) + \bb b(p)\inv \kappa(x)| + |\bb b(p)\inv \ell^r \e_\sigma| \\&\quad   \hspace{5em}+|\kappa(\Lambda(\ell)\ell\inv\e_x+x)-\kappa(x))|\\
&\le \max_{p\in\Omega_0}|\bb c(p) + \bb b(p)\inv \kappa(x)| + \mu\inv | \ell^r \e_\sigma|  +\rho_\kappa(|\hat x-x|)\\
&\le \satlev,
\end{align*}
in which we let $c(p)=-b(p)\inv q(p)$ and we used \eqref{pf:epsilon} and the fact that $|\bb b(p)\inv|\le \mu\inv$ for all $p$ (see Remark \ref{rmk:sat}). Hence, for all $t\in[T,\bar t)$, the control $u=\u(\hat{x},\hat{\sigma})$ is out of the saturation, and  similar arguments show that $\u(x,\hat{\sigma})$ is too. Thus, for all $t\in[T,\bar t)$, \eqref{pf:phi_u} and \eqref{pf:phi_sigma_lemma} yield
\begin{equation}\label{pf:dotx_bound} 
q(p) + b(p)u  =  \kappa(x) + \delta, 
\end{equation}
where 
\begin{equation}
\label{eq_def_delta}
\delta := -\ell^r\e_\sigma  + b(p)\bb\inv  (\kappa(\Lambda(\ell)\ell\inv e_x+x)-\kappa(x))
\end{equation}
which, in view of \eqref{pf:epsilon}, satisfies $|\delta(t)|\le \bar\delta$ in $[T,\bar t)$.
Since $p(T)\in \closed\ball_{\varrho_1}^\cB$, we conclude by definition of $\Omega_0$ in Section~\ref{sec:stabilizer}, that for every trajectory of the closed-loop system \eqref{s:cl} originating in $\Xb_0$, $(w,z,x,\eta,\hat x,\hat \sigma)$ is bounded, defined on $\Rplus$,  and such that $(w(t),z(t),x(t))\in\Omega_0$ for all $t\in\Rplus$. Thus, the first claim of the theorem holds.

\subsection{Practical Regulation}
Equations \eqref{pf:dotx_bound}-\eqref{eq_def_delta} and $p\in\Omega_0$ imply $|q(p) + b(p)u-\kappa(x)|\le \rho_x(\max\{|\sigma-\hat\sigma|,\ |x-\hat x| \})$ for all $t\ge T$ and for some $\rho_x\in\cK$, so that \eqref{e:thm_claim_21} follows by Lemma~\ref{lem:ell_epsilon} by noticing that $T$ can be taken arbitrarily small.

Regarding \eqref{e:thm_claim_22}, we observe that \eqref{pf:phi_u} and \eqref{pf:phi_sigma_lemma} imply 	
\begin{align*}
	|u-u\sr(w)| &\le  \big| c(p)-u\sr(w) +   b(p)\inv (\kappa(x)-\ell^r\e_\sigma)\\&\hspace{4em}
	+\bb\inv (\kappa(\Lambda(\ell)\ell\inv e_x+x)-\kappa(x))\big| \\
	&\le \rho_u \max\{ |p|_\cB ,\ |\hat\sigma-\sigma|,\ |\hat x-x|  \}   \\
	|\sigma+\bb u\sr(w)|&\le \rho_\sigma \max\{ |p|_\cB ,\ |\hat\sigma-\sigma|,\ |\hat x-x|  \}  
\end{align*} 
for   $t\ge T$ and for some  $\rho_u,\ \rho_\sigma >0$. Therefore, Lemma~\ref{lem:SICON}  implies that, for some $\rho_\cO\in\cK$,  $\limsup_{t\to\infty}|\xb(t)|_\cO \le\rho_\cO(\limsup_{t\to\infty}\max\{|\sigma(t)-\hat\sigma(t)|,\ |x(t)-\hat x(t)|\})$, so that  the second claim follows by Lemma \ref{lem:ell_epsilon}.

\subsection{Asymptotic Behavior}
We can write
\begin{equation*}
\begin{aligned}
\eta &= \tau(w) + \din,& u&= u\sr(w) + \dout 
\end{aligned}
\end{equation*}
with $u\sr$ given by \eqref{d:ustar} and
\begin{equation}\label{pf:id_dist}
\begin{aligned}
\din &:= \eta-\tau(w), & \dout &:= u-u\sr(w)
\end{aligned}
\end{equation}
that are bounded. Hence, if $(\cM,\Xi,\vhi,\Theta,\Thmap)$ satisfies the identifier requirement,  also the identifier has complete and bounded solutions.
Pick a solution $\xb=(w,z,x,\eta,\hat x,\hat \sigma,\xi,\theta)$ to \eqref{s:cl}, let $(\xi\sr,\theta\sr):\dom\xb\to\Xi\x\Theta$ be the trajectory produced by the identifier requirement for $(\din,\dout)$ given by \eqref{pf:id_dist}, and let $j\sr\in\N$ be such that $\xi\sr(j)\in\Xi\sr$ for all $j\ge j\sr$.
With $\vep\sr$ the optimal prediction error defined in \eqref{d:varepsilon_star}, let   for brevity $\vep\sr := \vep(\theta\sr,w)$ and change variables as $\e\mapsto \zeta$, where
\begin{equation*}
\zeta := (\zeta_x,\,\zeta_\sigma) = (\e_x,\, \e_\sigma -\ell^{-r} b(p) \vep\sr) .
\end{equation*}
In view of \eqref{pf:dot_e_x}, $\zeta_x$ jumps as $\zeta_x^+=\zeta_x$ and flows according to 
\begin{equation}\label{pf:dot_zeta_x}
\dot\zeta_x = \ell(A-HC)\zeta_x + \ell B( \zeta_\sigma + \ell^{-r} b(p)\vep\sr  ) +B \Delta_{3,\ell}
\end{equation}
with $|\Delta_{3,\ell}|\leq a_2|\zeta_x|$.
In view of the stability analysis of Section \ref{sec:proof:stability}, if $\ell>\ell\sr_{\rm s}$ then for all $t\ge T$, $\sigma$ assumes the expression \eqref{pf:phi_sigma_lemma}, so that the quantity $\sigma\sr:=\sigma+b(p)\vep\sr$ satisfies $\zeta_\sigma=\ell^{-r}(\hat\sigma-\sigma\sr)$ and, for all $t\ge T$,
\begin{equation}\label{pf:d:sigma_star}
\sigma\sr = -\bb c(p)+\bb \vep\sr + \overline{m}(p)\big(\kappa(x)-\ell^r\zeta_\sigma\big)
\end{equation}
where we let
$
\overline{m}(p) := I-\bb b(p)\inv
$. 
Since
\begin{align*}
 \dot\vep\sr =   L_{s(w)}u\sr(w) - 
  \pd{\hat\gamma}{\eta}(\theta\sr,\tau(w))\, (F\tau(w)+Gu\sr(w))
\end{align*}
and,  by definition of $\psi$,  
\begin{equation*}
\psi(\theta\sr,\tau(w),u\sr(w)) = 
\pd{\hat \gamma}{\eta}(\theta\sr,\tau(w)) \,(F\tau(w)+Gu\sr(w)),
\end{equation*} 
for all $(t,j)\succeq(T,j\sr)$,  then 
\begin{equation*}
\dot\sigma\sr =-\bb \psi(\theta\sr,\tau(w),u\sr(w)) +K_1  + K_{2,\ell} + K_{3,\ell} - \overline{m}(p) \ell^r \dot\zeta_\sigma 
\end{equation*}
for all $(t,j)\succeq (T,j\sr)$, where
\begin{align*}
K_1   &:= \bb\left( L_{s(w)}u\sr(w) - L_{\fb(p,u)}c(p) \right)\\
K_{2,\ell}  &:= \bb b(p)\inv (L_{\fb(p,u)}b(p)) b(p)\inv \big(\kappa(x)-\ell^r\zeta_\sigma\big) \\
K_{3,\ell}  &:= \overline{m}(p)  \kappa'(x) (\kappa(x) +\delta) 
\end{align*}
with $\delta$ defined in \eqref{eq_def_delta}.
Thus, $\zeta_\sigma$ jumps according to
\begin{equation}\label{pf:zeta_sigma_plus}
\zeta_\sigma^+ = \zeta_\sigma + \ell^{-r}b(p)(\vep\sr-\vep\sr{}^+)
\end{equation}
in which we let $\vep\sr{}^+:=\vep(\Thmap(\xi\sr),w)$, and,   for all $(t,j)\succeq(T_,j\sr)$,   it flows according to
\begin{equation}\label{pf:dot_zeta_sigma0}
\begin{aligned}
\dot\zeta_\sigma  & =  -\ell H_{r+1}C\zeta_x   +\overline m(p)\dot\zeta_\sigma    +\ell^{-r}   K
\end{aligned}
\end{equation}
in which
\begin{equation*}
K:=-K_1 -K_{2,\ell} -K_{3,\ell} - \bb \big( \psi(\theta,\eta,u)-\psi(\theta\sr,\tau(w),u\sr(w))  \big).
\end{equation*}
Notice that $I-\overline{m}(p) = \bb b(p)\inv$ is uniformly invertible and bounded (see Remark \ref{rmk:sat}). Then, solving \eqref{pf:dot_zeta_sigma0} for $\dot\zeta_\sigma$ yields
\begin{equation}\label{pf:dot_zeta_sigma}
\dot\zeta_\sigma = -\ell H_{r+1} C\zeta_x - (b(p)-\bb)\bb\inv \ell H_{r+1} C\zeta_x + \ell^{-r} b(p)\bb\inv K.
\end{equation}
Hence, $\zeta$ satisfies $\zeta^+ = \zeta + \Be_\sigma \ell^{-r} b(p)(\vep\sr-\vep\sr{}^+)$ 
during jumps and, in view of \eqref{pf:dot_zeta_x} and \eqref{pf:dot_zeta_sigma},
\begin{multline*}
\dot\zeta = \ell \Ae \zeta +\Be_\sigma\big( -\alpha \ell H_{r+1}C\zeta_x + \ell^{-r}b(p)\bb\inv K   \big) 
\\
+ \Be_x(\ell^{1-r}b(p)\vep\sr +\Delta_{3,\ell} )
\end{multline*} 
during flows, with $\alpha:= (b(p)-\bb)\bb\inv$ that due to A\ref{ass:b} satisfies $|\alpha|\le 1-\mu$ everywhere.
In view of A\ref{ass:regularity} and since $p\in\Omega_0$ for all $t\in\Rplus$, there exist $\nu_1,\nu_2,\nu_3>0$ such that
\begin{equation*}
\begin{aligned}
|b(p)\bb\inv K_1|&\le \nu_1\big( |p|_\cB + |u-u\sr(w)|  \big),\\
|b(p)\bb\inv K_{2,\ell}|&\le \nu_2\big( |p|_\cB  + \ell^{r }|\zeta|\big),\\
|b(p)\bb\inv K_{3,\ell}|&\le \nu_3\big(|p|_\cB+ \ell^r|\zeta| + |\vep\sr| \big)
\end{aligned}
\end{equation*}
for  $t\ge T$.
Regarding the term $\psi(\theta,\eta,u)-\psi(\theta\sr,\tau(w),u\sr(w))$, we  notice that for all $j\ge j\sr$, $\theta\sr(j)\in\Theta\sr=\Thmap(\Xi\sr)$, while $(\tau(w),u\sr(w))\in\cH\sr\x\cU\sr$ holds everywhere by construction. Thus, since in view of the identifier requirement $\partial\hat\gamma/\partial\eta$ is locally Lipschitz and $|\theta-\theta\sr|\le \rho_\theta(|\xi-\xi\sr|)$ with $\rho_\theta$ locally Lipschitz for all $(\theta,\theta\sr,\xi,\xi\sr)\in \Theta\x\Theta\sr\x\Xi\x\Xi\sr$, and since $\psi$ is globally bounded,  there exists $\nu_4>0$ such that  
\begin{equation*}
\begin{aligned}
|\psi(\theta&,\eta,u)-\psi(\theta\sr,\tau(w),u\sr(w))|\\&\le \nu_4 \big( |\xi-\xi\sr| + |\eta-\tau(w)|  + |u-u\sr(w)| \big)
\end{aligned}
\end{equation*} 
 holds for each $(t,j)\succeq (T,j\sr)$. 
In view of \eqref{pf:phi_u} and \eqref{pf:d:sigma_star}, and since for $t\ge T$  the control is out of saturation, then
\begin{equation}\label{pf:bound_u_usr}
|u-u\sr(w) | \le \nu_5\big( |p|_\cB + |\vep\sr| + \ell^r|\zeta|\big)
\end{equation}
for some $\nu_5>0$.
Therefore, the last term of \eqref{pf:dot_zeta_sigma} satisfies
\begin{equation*}
|\ell^{-r} b(p)\bb\inv  K|\le \nu_6  |\zeta|   + \ell^{-r}\nu_7\big( |(p,\eta)|_\cE + |\xi-\xi\sr|+|\vep\sr| \big)
\end{equation*}
for all $(t,j)\succeq(T,j\sr)$, 
with $\nu_6:=\max\{ \nu_2+\nu_3,\,(\nu_1+\nu_4)\nu_5 \}$ and $\nu_7:=\max\{ \nu_1+\nu_2+\nu_3,\, (\nu_1+\nu_4)\nu_5,\, \nu_3,\,\nu_4 \}$ and with $\cE$  given by \eqref{d:cE}.  Hence, Lemma \ref{lem:e_mu} and \eqref{pf:zeta_sigma_plus} yield the existence of constants $\nu_8,\nu_9,\nu_{10},\nu_{11}>0$ such that  
\begin{equation}\label{pf:iss_zeta}
\begin{aligned}
&|\zeta(t,j)|\le \max\Big\{ \nu_8\ee^{-\ell\nu_9 (t-t_j)}|\zeta(t_j,j)|,\,\nu_{10}\ell^{-r}|\vep\sr|_{(t,j)}\\&\hspace{7em} \nu_{10}\ell^{-(r+1)} |(p,\eta)|_{\cE,t} ,\,\nu_{10}\ell^{-(r+1)}|\xi-\xi\sr|_j \Big\} \\
&|\zeta(t^j,j+1)|\le 2\max\Big\{|\zeta(t^j,j)|,\\&\hspace{10em}\nu_{11}\ell^{-r}|(\vep\sr(t^j,j),\vep\sr{}^+(t^j,j))|\Big\}
\end{aligned}
\end{equation}
for all $(t,j)\succeq (T,j\sr)$. We then observe that, since for each $(t,j)\in\dom\xb$, $t^j-t_j\ge\unT$, then for each constant $\bar \nu>0$, sufficiently large values of $\ell$ yield
\begin{equation*}
\lim_{{t+j\to\infty,\,(t,j)\in\dom\xb}} \bar \nu^{j+1} \ee^{-\ell\nu_9 t} \le  \lim_{ j\to\infty } \bar \nu^{j+1} \ee^{-\ell\nu_9 j\unT }  = 0.
\end{equation*}
Thus, there follows from \eqref{pf:iss_zeta} by induction and standard ISS arguments that there exist $\nu_{12}>0$ and $\ell_{3}\sr(\unT)>\ell_{\rm s}$ such that  $\ell>\ell_{3}\sr(\unT)$ implies
\begin{equation}\label{pf:ls_zeta}
\begin{aligned}
\limsup_{t\to\infty }|\zeta(t)| \le& \nu_{12} \ell^{-(r+1)} \max\Big\{
\limsup_{t\to\infty}|(p(t),\eta(t))|_\cE,\\
& \limsup_{j\to\infty}|\xi(j)-\xi\sr(j)|,\, \ell \limsup_{t+j\to\infty}|\vep\sr(t,j)|
\Big\},
\end{aligned}
\end{equation}
in which we used the fact that $\limsup_{t+j\to\infty }|\vep\sr(t,j)|=\limsup_{t+j\to\infty }|\vep\sr{}^+(t,j)|$.
We now observe that \eqref{pf:bound_u_usr} implies that the flow equation of $(w,z,\eta)$ can be written as \eqref{s:zerodyn_eta}, with $\delta_1+\delta_2=u-u\sr(w)$ such that $|\delta_1|\le \nu_5|p|_\cB$ and $|\delta_2|\le \nu_5(|\vep\sr|+\ell^r|\zeta|)$. Moreover, substituting \eqref{pf:phi_u}, \eqref{pf:d:sigma_star} into the equation of $x$ yields \eqref{s:x_kappa} 
for all $t\ge  T$, with $\delta$ defined in \eqref{eq_def_delta} that, for some $\nu_{13}>0$, fulfills $|\delta|\le \nu_{13}(|\vep\sr|+\ell^r|\zeta|)$. 
Hence,  Lemma \ref{lem:SICON} and the ISS property of \eqref{s:x_kappa} yield the existence of  $\nu_{12}>0$  such that
\begin{equation}\label{pf:ls_cE}
\begin{aligned}
\limsup_{t\to\infty} |(p(t),\eta(t))|_\cE \le \nu_{13}\max\Big\{
&\limsup_{t+j\to\infty}|\vep\sr(t,j)|,\\
&\ell^r\limsup_{t\to\infty}|\zeta(t)|
\Big\}
\end{aligned}
\end{equation}
where we recall that $\cE$ is given by \eqref{d:cE}.
%
%
On the other hand, the identifier requirement, the expression \eqref{pf:id_dist} and the bounds above yield the existence of a $\nu_{14}>0$ such that
\begin{equation} \label{pf:ls_xi}
\begin{aligned}
&\limsup_{j\to\infty }|\xi(j)-\xi\sr(j)| \le \nu_{14}\max\Big\{  \limsup_{t+j\to\infty }|\vep\sr(t,j)| , \\
&\qquad\qquad\qquad\limsup_{t\to\infty}|(p(t),\eta(t))|_\cE ,\,   \ell^r\limsup_{t\to\infty }|\zeta(t)| \Big\}
\end{aligned}
\end{equation}
 Denote   $|(p,\eta,\xi)|_{\cE\sr}:=\max\{ |(p,\eta)|_\cE,\middlebreak|\xi-\xi\sr| \}$. 
Then, with $\nu_{15}:=\max\{\nu_{13},\nu_{14},\nu_{13}\nu_{14}\}$,    \eqref{pf:ls_cE} and \eqref{pf:ls_xi} yield
 \begin{equation}\label{pf:ls_cEsr}
 \begin{aligned}
& \limsup_{t+j\to\infty}|(p(t),\eta(t),\xi(j))|_{\cE\sr} \\&\quad \le \nu_{15} \max\Big\{
\ell^r  \limsup_{t\to\infty}|\zeta(t)|, 
 \limsup_{t+j\to\infty}|\vep\sr(t,j)|
 \Big\}.
 \end{aligned}
 \end{equation}
With  $\cO$ and  $|\cdot|_{\cO\sr}$ defined respectively in \eqref{d:cO} and \eqref{d:dist_cOsr}, we observe that $|\xb|_{\cO\sr} \le 
\nu_{16}  \max\{ |(p,\eta,\xi)|_{\cE\sr} ,\, \ell^r|\zeta| \}$ 
for some  $\nu_{16}>0$. 
Therefore,  
substituting \eqref{pf:ls_zeta} into \eqref{pf:ls_cEsr}   and using $|\zeta|\le|\xb|_{\cO\sr}$ and $|(p,\eta,\xi)|_{\cE\sr}\le|\xb|_{\cO\sr}$ yields
\begin{equation*}
\begin{aligned}
&\limsup_{t+j\to\infty }|\xb(t,j)|_{\cO\sr}\\& \le \ \nu_{17}\max\Big\{  \ell\inv \limsup_{t+j\to\infty}|\xb(t,j)|_{\cO\sr} ,   \limsup_{t+j\to\infty}|\vep\sr(t,j)| \Big\}.
\end{aligned}
\end{equation*}
with $\nu_{17}:= \nu_{16}\max\{\nu_{12},\nu_{15},\nu_{12}\nu_{15} \}$, and the claim follows with $\alpha_\xb=\nu_{17}$ by taking $\ell_{\vep}\sr(\unT) >\max\{ \ell\sr_{3}(\unT),\, \nu_{17}\}$.
\hfill \QED

%
%

\section{Conclusion}
In this paper we proposed a regulator design for a class of multivariable nonlinear systems which employs an adaptive internal model unit and an extended high-gain observer to solve instances of practical, approximate and asymptotic output regulation problems. The proposed design employs system identification algorithms to carry out the estimation of an optimal internal model, and does not rely on high-gain stabilization techniques. Future research directions will be aimed at exploiting the additional freedom on the stabilizer to deal with non minimum-phase systems, and at investigating further identification algorithms that fits in the framework, thus developing further the bridge with the system identification literature. 
We also aim to study the robustness of the proposed scheme in the formal framework of \cite{Bin2018robustness}, by connecting the identifier's validation to classical robustness concepts.

%
%
%
%

\appendix

\subsection{Proof of Proposition \ref{prop:minibatch}}\label{apd:proof_minibatch}
Consider the interconnection \eqref{s:core_identifier} with $(\Xi,\vhi,\Theta,\Thmap)$ given in Section\ref{sec:minibatch} and $(\win,\wout)=(\tau(w),u\sr(w))$. Define $\xi_1\sr(j) =\col(\tau_1^j,\dots,\tau_N^j)$ and $\xi_2\sr(j)=\col(u\sr_1{}^j,\dots, u\sr_N{}^j)$ in which
\begin{align*}
\tau_i^j &= \begin{cases}
0 & \text{if } j+i-N-1<0\\
\tau(w(t^{j+i-N-1})) &\text{otherwise}
\end{cases} \\
u\sr_i{}^j &= \begin{cases}
0 & \text{if } j+i-N-1<0\\
u\sr(w(t^{j+i-N-1})) &\text{otherwise}
\end{cases}
\end{align*}
and let $\theta\sr(j):=\cG(\lambda_{\deta}(\xi_1\sr(j)),\lambda_{\dy}(\xi_2\sr(j)))$. Then, in view of \eqref{s:mb_xi}, for each solution pair $((\tim,w,\xi),(\din,\dout))$ to \eqref{s:core_identifier}, $((\tim,w,\xi\sr),(0,0))$ is a solution pair to \eqref{s:core_identifier}. Moreover, for $j\ge j\sr:= N$, $\cJ_{(\tim,w)}(j)(\theta) = \cI^N_{(s_{\rm in}^j,s_{\rm out}^j)}(\theta)$ with $s_{\rm in}^j:=\lambda_\deta(\xi_1\sr(j))$ and $s_{\rm out}^j=\lambda_\dy(\xi_2\sr(j))$. Hence the optimality and regularity items of the identifier requirements follow by A\ref{ass:minibatch}.   Finally, the   stability item follows by   the fact that the system $\tilde \xi:=\xi-\xi\sr$ is an asymptotically stable linear system driven by the input $(\din,\dout)$, and hence it is ISS relative to the origin and with respect to the input $(\din,\dout)$ with linear gains. \hfill\QED

 
\subsection{Proof of Lemma \ref{lem:e_mu}} \label{apd:proof_lemma_e_mu}
The proof follows by the same arguments of \cite{Freidovich2008}. In particular,  we first consider the system
\begin{equation}\label{pfl:chi}
\dot\chi = \ell\Ae\chi + \ell \Be_\sigma H_{r+1}  \delta_0 + \Be_x\delta_1 + \delta_2 
\end{equation}
with $|\delta_1|\le \pi_1|\chi|$. Since $\Ae$ is Hurwitz there exists $P=P^\top>0$ fulfilling $\Ae^\top P+P\Ae=- I$ and such that the Lyapunov candidate $V(\chi):=\sqrt{\chi^\top P\chi}$ satisfies $\und\lambda|\chi|\le V(\chi)\le \bar\lambda|\chi|$, with $\und\lambda$ and $\bar\lambda$ respectively the smallest and largest eigenvalues of $P$. Then, there exist   $b_1,b_2,b_3>0$ such that, for all $\chi\ne 0$,
\begin{equation*}
\begin{aligned}
\dot V(\chi) &\le -b_1 \ell V(\chi)   +  b_2(\ell |\delta_0| + |\delta_1|+|\delta_2|)\\
&\le -(b_1\ell-b_3)V(\chi) + b_2(\ell|\delta_0|+|\delta_2|)
\end{aligned}
\end{equation*}
Let $\ell_0\sr:= 2 b_3/b_1$. Then $b_3-b_1\ell\le -b_1\ell/2$ for all $\ell\ge\ell\sr_0$, and this shows that $\chi$ is ISS relative to the origin and with respect to the inputs $\delta_0$ and $\delta_2$. Moreover, the asymptotic gain between $\delta_2$ and $|\chi|$ is of the form $b_4/\ell$, for some $b_4$ independent on $\ell$.

Regarding the asymptotic gain between $\delta_0$ and $|\chi|$, we observe that \eqref{pfl:chi} is a linear system, and there follows from the structure of $\Ae$, $\Be$ and $C$, that when $\delta_1=0$ and $\delta_2=0$, 
\[
C\chi_1^{(r+1)} + H_1 C\chi_1^{(r)}+\cdots + H_r C\dot\chi_1 + H_{r+1}C\chi_1 = H_{r+1}  \delta_0.
\]
Thus, the transfer function from $\delta_0$ to $C\chi_1$ has the form
\begin{equation*}
(s^{r+1}I + H_1 s^{r} + \cdots + H_r s + H_{r+1})\inv H_{r+1}.
\end{equation*}
Since all the poles are real and negative by construction, and since each $H_i$ is diagonal, it follows that the gain between $\delta_0$ and $|C\chi_1|$ is unitary. Finally the proof follows by standard small-gain arguments by observing that system \eqref{pf:chi} is obtained as the interconnection of \eqref{pfl:chi} and the algebraic system $\delta_0 =-\alpha C\chi_1$  and that, since $|\alpha|\le\bar\alpha<1$ the overall gain is less than one.\hfill\QED

\subsection{Proof of Lemma \ref{lem:ell_epsilon}} \label{apd:proof_lemma_ell_epsilon}
	In view of \eqref{pf:e}, if $\ell\ge\ell\sr_0$ \eqref{pf:e_bound} leads to the existence of $\pi_1,\pi_2>0$ such that
	\begin{equation*}
	\begin{aligned}
	\max&\{|x(t)-\hat x(t)|,|\sigma(t)-\hat\sigma(t)|\} \\&\le \max\Big\{ \ell^r \pi_1 e^{-a_4\ell t} \max\{|x(0)-\hat x(0)|,|\sigma(0)-\hat\sigma(0)|\},\,\\&\qquad\qquad \pi_2 \ell\inv |\Delta_5|_t   \Big\}.
	\end{aligned}
	\end{equation*}
	As $\Xb_0$ is compact, there exists $b>0$ such that $\max \{|x(t)-\hat x(t)|,|\sigma(t)-\hat\sigma(t)|\} \le \max\{b  \ell^r\exp(-\ell a_4 t), \pi_2 a_1 \ell\inv   \}$ for all $t\in(0,T_0)$. Pick $T\in(0,T_0)$ and $\epsilon>0$ arbitrarily, and let
	\begin{equation*}
	\bar t(\ell,\epsilon) := \dfrac{r}{\ell a_4} \log\left(\ell \sqrt[r]{b/\epsilon}\right).
	\end{equation*}
	Then  $	\lim_{\ell\to\infty} \bar t(\ell,\epsilon) = 0$, 
	so that there exists $\bar\ell(\epsilon,T)>0$ such that, for all $\ell\ge\bar\ell(\epsilon,T)$, $\bar t(\ell,\epsilon)\le T$, and $t\ge \bar t(\ell,\epsilon)$ yields
$b\ell^r \exp(-\ell a_4 t)  \le b\ell^r \exp \left( -\log\left(\ell^r b/\epsilon \right) \right) = \epsilon$. 
	Hence, the claim holds with   $\ell\sr_1(T,\epsilon) :=\max \{ \ell\sr_0,\, \bar\ell(\epsilon,T),\,  \pi_2 a_1/\epsilon \}$. \hfill\QED

\subsection{Proof of Lemma \ref{lem:sigma}}\label{apd:proof_lemma_sigma}

We want to show that $\phi_\sigma(p,\e_\sigma)=\phi$ with $\phi$ the quantity defined by
$
\phi:=\bb b(p)\inv q(p) +  (I-\bb b(p)\inv) \big( \kappa(x)-\ell^r\e_\sigma \big) 
$.
	First, notice that the quantity $s:=-\phi -\ell^r\e_\sigma + \kappa(x)$ satisfies
	\begin{equation*}
	s =  -\bb b(p)\inv q(p)  + \bb b(p)\inv\big(\kappa(x) - \ell^r\e_\sigma \big).
	\end{equation*}
In view of Remark \ref{rmk:sat}, $|\bb b(p)\inv|\le\mu\inv$ for all $p$. If $p\in\Omega_0$ and $|\ell^r\e_\sigma|\le \mu \varrho_2$,  in turn,
	we get
	\begin{align*}
	|s|&\le \max_{p\in\Omega_0}|\bb c(p) + \bb b(p)^{-1}\kappa(x)| + \mu\inv|\ell^r\e_\sigma| 
	\\
	&\le  \max_{p\in\Omega_0}|\bb c(p) + \bb b(p)^{-1}\kappa(x)| + \varrho_2 \le \satlev.
	\end{align*}
	Hence, $\sat(s)=s$, and it is easy to see that  $T_\sigma(p,\e_\sigma,\phi)=~0$. By uniqueness of solutions of \eqref{pf:eq_Tsigma}, we conclude $\phi=\phi_\sigma(p,e_\sigma)$ which is the claim.
	\hfill \QED


\bibliographystyle{ieeetran}
\bibliography{biblio}

\begin{thebibliography}{10}
\providecommand{\url}[1]{#1}
\csname url@samestyle\endcsname
\providecommand{\newblock}{\relax}
\providecommand{\bibinfo}[2]{#2}
\providecommand{\BIBentrySTDinterwordspacing}{\spaceskip=0pt\relax}
\providecommand{\BIBentryALTinterwordstretchfactor}{4}
\providecommand{\BIBentryALTinterwordspacing}{\spaceskip=\fontdimen2\font plus
\BIBentryALTinterwordstretchfactor\fontdimen3\font minus
  \fontdimen4\font\relax}
\providecommand{\BIBforeignlanguage}[2]{{%
\expandafter\ifx\csname l@#1\endcsname\relax
\typeout{** WARNING: IEEEtran.bst: No hyphenation pattern has been}%
\typeout{** loaded for the language `#1'. Using the pattern for}%
\typeout{** the default language instead.}%
\else
\language=\csname l@#1\endcsname
\fi
#2}}
\providecommand{\BIBdecl}{\relax}
\BIBdecl

\bibitem{Francis1976}
B.~A. Francis and W.~M. Wonham, ``The internal model principle of control
  theory,'' \emph{Automatica}, vol.~12, pp. 457--465, 1976.

\bibitem{Davison1976}
E.~J. Davison, ``The robust control of a servomechanism problem for linear
  time-invariant multivariable systems,'' \emph{IEEE Trans. Autom. Contr.},
  vol. AC-21, no.~1, pp. 25--34, 1976.

\bibitem{Huang1990}
J.~Huang and W.~J. Rugh, ``On a nonlinear multivariable servomechanism
  problem,'' \emph{Automatica}, vol.~26, no.~6, pp. 963--972, 1990.

\bibitem{Isidori1990}
A.~Isidori and C.~I. Byrnes, ``Output regulation of nonlinear systems,''
  \emph{IEEE Trans. Autom. Contr.}, vol.~35, no.~2, pp. 131--140, 1990.

\bibitem{Huang1995}
J.~Huang, ``Asymptotic tracking and disturbance rejection in uncertain
  nonlinear systems,'' \emph{IEEE Trans. Autom. Contr.}, vol.~40, no.~6, pp.
  1118--1122, 1995.

\bibitem{Byrnes1997}
C.~Byrnes, F.~{Delli Priscoli}, and A.~Isidori, ``Structurally stable output
  regulation for nonlinear systems,'' \emph{Automatica}, vol.~33, no.~3, pp.
  369--385, 1997.

\bibitem{Byrnes2003}
C.~I. Byrnes and A.~Isidori, ``Limit sets, zero dynamics and internal models in
  the problem of nonlinear output regulation,'' \emph{IEEE Trans. Autom.
  Contr.}, vol.~48, pp. 1712--1723, Oct. 2003.

\bibitem{Byrnes2004}
------, ``Nonlinear internal models for output regulation,'' \emph{IEEE Trans.
  Autom. Contr.}, vol.~49, pp. 2244--2247, Dec. 2004.

\bibitem{Marconi2007}
L.~Marconi, L.~Praly, and A.~Isidori, ``Output stabilization via nonlinear
  {L}uenberger observers,'' \emph{SIAM J. Contr. Opt.}, vol.~45, pp.
  2277--2298, 2007.

\bibitem{Byrnes2003b}
C.~I. Byrnes, A.~Isidori, and A.~Praly, ``On the asymptotic properties of a
  system arising from the non-equilibrium theory of output regulation,''
  \emph{Mittag Leffler Institute}, 2003.

\bibitem{Wang2016}
L.~Wang, A.~Isidori, H.~Su, and L.~Marconi, ``Nonlinear output regulation for
  invertible nonlinear {MIMO} systems,'' \emph{Int. J. Robust Nonlinear
  Control}, vol.~26, pp. 2401--2417, 2016.

\bibitem{WanIsiLiuSu}
L.~Wang, A.~Isidori, Z.~Liu, and H.~Su, ``Robust output regulation for
  invertible nonlinear {MIMO} systems,'' \emph{Automatica}, vol.~82, pp.
  278--286, 2017.

\bibitem{Bin2019}
M.~Bin and L.~Marconi, ``Output regulation by postprocessing internal models
  for a class of multivariable nonlinear systems,'' \emph{Int. J. Robust
  Nonlinear Control}, vol.~3, pp. 1115--1140, 2020.

\bibitem{Bin2018robustness}
M.~Bin, D.~Astolfi, L.~Marconi, and L.~Praly, ``About robustness of internal
  model-based control for linear and nonlinear systems,'' in \emph{57th IEEE
  Conference on Decision and Control}, 2018.

\bibitem{Marconi2008}
L.~Marconi and L.~Praly, ``Uniform practical nonlinear output regulation,''
  \emph{IEEE Trans. Autom. Contr.}, vol.~53, pp. 1184--1202, 2008.

\bibitem{Astolfi2015}
D.~Astolfi, L.~Praly, and L.~Marconi, ``Approximate regulation for nonlinear
  systems in presence of periodic disturbances,'' in \emph{54th IEEE Conference
  on Decision and Control}, 2015, pp. 7665--7670.

\bibitem{Isidori2012}
A.~Isidori, L.~Marconi, and L.~Praly, ``Robust design of nonlinear internal
  models without adaptation,'' \emph{Automatica}, vol.~48, pp. 2409--2419,
  2012.

\bibitem{Freidovich2008}
L.~B. Freidovich and H.~K. Khalil, ``Preformance recovery of
  feedback-linearization-based designs,'' \emph{IEEE Trans. Autom. Contr.},
  vol.~53, no.~10, pp. 2324--2334, 2008.

\bibitem{Serrani2001}
A.~Serrani, A.~Isidori, and L.~Marconi, ``Semiglobal nonlinear output
  regulation with adaptive internal model,'' \emph{IEEE Trans. Autom. Control},
  vol.~46, no.~8, pp. 1178--1194, 2001.

\bibitem{DelliPriscoli2006}
F.~D. Priscoli, L.~Marconi, and A.~Isidori, ``A new approach to adaptive
  nonlinear regulation,'' \emph{{SIAM} J. {C}ontrol {O}ptim.}, vol.~45, no.~3,
  pp. 829--855, 2006.

\bibitem{Pyrkin2018}
A.~Pyrkin and A.~Isidori, ``Output regulation for robustly minimum-phase
  multivariable nonlinear systems,'' in \emph{IEEE 56th Conference on Decision
  and Control}, 2017, pp. 873--878.

\bibitem{Bin2018d}
M.~Bin, L.~Marconi, and A.~R. Teel, ``Adaptive output regulation for linear
  systems via discrete-time identifiers,'' \emph{Automatica}, vol. 105, pp.
  422--432, 2019.

\bibitem{Forte2017}
F.~Forte, L.~Marconi, and A.~R. Teel, ``Robust nonlinear regulation:
  Continuous-time internal models and hybrid identifiers,'' \emph{IEEE Trans.
  Autom. Contr.}, vol.~62, no.~7, pp. 3136--3151, 2017.

\bibitem{Bin2018}
M.~{Bin} and L.~{Marconi}, ````{C}lass-type'' identification-based internal
  models in multivariable nonlinear output regulation,'' \emph{IEEE Trans.
  Autom. Control}, to appear.

\bibitem{BinBerMar}
P.~Bernard, M.~Bin, and L.~Marconi, ``Adaptive output regulation via nonlinear
  luenberger observer-based internal models and continuous-time identifiers,''
  \emph{Automatica}, to appear.

\bibitem{Li2014}
S.~Li, J.~Yang, W.-H. Chen, and X.~Chen, \emph{Disturbance Observer-Based
  Control: Methods and Applications}.\hskip 1em plus 0.5em minus 0.4em\relax
  CRC Press, 2014.

\bibitem{Goebel2012book}
R.~Goebel, R.~G. Sanfelice, and A.~R. Teel, \emph{Hybrid Dynamical
  Systems}.\hskip 1em plus 0.5em minus 0.4em\relax Princeton, N. J.: Princeton
  University Press, 2012.

\bibitem{Ljung1999}
L.~Ljung, \emph{System Identification: Theory for the user}.\hskip 1em plus
  0.5em minus 0.4em\relax Prentice Hall, 1999.

\bibitem{Khalil1998}
H.~K. Khalil, ``On the design of robust servomechanisms for minimum phase
  nonlinear systems,'' in \emph{37th IEEE Conference on Decision and Control},
  1998, pp. 3075--3080.

\bibitem{CampbellPseudoInvBook}
S.~Campbell and C.~Meyer, \emph{Generalized Inverses of Linear
  Transformations}.\hskip 1em plus 0.5em minus 0.4em\relax SIAM, 2009.

\bibitem{Sjoberg1995}
J.~Sj{\"o}berg, Q.~Zhang, L.Ljung, A.~Benveniste, B.~Delyon, P.~Glorennec,
  H.~Hjalmarsson, and A.~Juditsky, ``Nonlinear black-box modeling in system
  identification: a unified overview,'' \emph{Automatica}, vol.~31, pp.
  1691--1724, 1995.

\bibitem{Andrieu2008}
V.~Andrieu, L.~Praly, and A.~Astolfi, ``Homogeneous approximation, recursive
  observer design, and output feedback,'' \emph{SIAM J. Contr. Opt.}, vol.~47,
  no.~4, pp. 1814--1850, 2008.

\bibitem{Sjoberg1993}
J.~Sj\"{o}berg, T.~McKelvey, and L.~Ljung, ``On the use of regularization in
  system identification,'' in \emph{IFAC 12th World Congress}, 1993.

\bibitem{LeeSmi}
R.~Lee and M.~Smith, ``Nonlinear control for robust rejection of periodic
  disturbances,'' \emph{Systems \& Control Letters}, vol.~39, pp. 97--107,
  2000.

\end{thebibliography}

\end{document}